\documentclass[preprint,showpacs,preprintnumbers,amsmath,amssymb,superscriptaddress,nofootinbib]{revtex4}
\usepackage{graphicx,color}
\usepackage{amsmath,amssymb}
\usepackage{url}
\usepackage{epstopdf}
\newcommand{\hs}{\hspace*{0.5cm}}

\newcommand{\be}{\begin{equation}}
\newcommand{\ee}{\end{equation}}
\newcommand{\bea}{\begin{eqnarray}}
\newcommand{\eea}{\end{eqnarray}}
\newcommand{\nn}{\nonumber}
\newcommand{\crn}{\nonumber \\}

\newcommand{\al}{\alpha}

\newcommand{\om}{\omega}

\newcommand{\fr}{\frac}

\newcommand{\bc}{\begin{center}}
\newcommand{\ec}{\end{center}}

\newcommand {\ba}{\begin{array}}
\newcommand {\ea}{\end{array}}
\newcommand{\ben}{\begin{enumerate}}
\newcommand{\een}{\end{enumerate}}

\usepackage{bm}
\usepackage{dcolumn}

\begin{document}

\title{ Contribution of heavy neutrinos to decay of standard-model-like Higgs boson $h\rightarrow \mu\tau$ in a 3-3-1 model with additional gauge singlets}

\author{H. T. Hung }\email{hathanhhung@hpu2.edu.vn (Corresponding Author)} 
\affiliation{Department of Physics, Hanoi Pedagogical University 2, Phuc Yen,  Vinh Phuc 280000, Vietnam}

\author{N.T. Tham}\email{nguyenthitham@hpu2.edu.vn}
\affiliation{Department of Physics, Hanoi Pedagogical University 2, Phuc Yen,  Vinh Phuc 280000, Vietnam}

\author{T.T. Hieu}\email{tranhieusp2@gmail.com}

\affiliation{Department of Physics, Hanoi Pedagogical University 2, Phuc Yen,  Vinh Phuc 280000, Vietnam}

\author{N.T.T. Hang}\email{hangntt@daihocpccc.edu.vn}

\affiliation{The University of fire prevention and fighting, 243 Khuat Duy Tien, Nhan Chinh, Thanh Xuan, Hanoi 100000, Vietnam}

\begin{abstract}
 In the framework of  the improved version of the 3-3-1 models with right-handed neutrinos, which is added to the Majorana neutrinos as new gauge singlets, the recent experimental neutrino oscillation  data is completely explained through the inverse seesaw mechanism.   We show that the major contributions to $Br(\mu\rightarrow e\gamma)$ are derived from corrections at 1-loop order of heavy neutrinos and bosons. But, these contributions are  sometimes mutually destructive, creating regions of parametric spaces where the experimental limits of $Br(\mu\rightarrow e\gamma)$ are satisfied. In these regions, we find that  $Br(\tau\rightarrow \mu\gamma)$ can achieve values of $10 ^{- 10}$ and $Br(\tau\rightarrow e\gamma)$ may even reach values of $10 ^{- 9}$ very close to the upper bound of the current experimental limits. Those are ideal areas to study lepton-flavor-violating decays of the standard- model- like Higgs boson ($h_1^0$). We also pointed out that the contributions of heavy neutrinos play an important role to change $Br(h_1^0\rightarrow \mu\tau)$, this is presented through different forms of mass mixing matrices ($M_R$) of heavy neutrinos. When $M_R \sim diag(1,1,1)$, $Br(h_1^0\rightarrow \mu\tau)$ can get a greater value than the cases $M_R \sim diag(1,2,3)$ and $M_R \sim diag(3,2,1)$ and the largest that $Br(h_1^0\rightarrow \mu\tau)$ can reach is very close $10 ^{-3}$.
\end{abstract} 
\maketitle
 \section{\label{intro} Introduction}
 \allowdisplaybreaks
 The current experimental data has demonstrated that the neutrinos are massive and  flavor oscillations \cite{Zyla:2020zbs}. This leads to a consequence that lepton-flavor-violating decays of
 charged leptons (cLFV) must exist and it is strongly dependent on the flavor neutrino oscillation. The cLFV is concerned in both theory and experiment. On the theoretical side, the processes $l_a\rightarrow l_b\gamma$ are loop induced, we therefore pay attention  to both neutrino and boson contributions, with special attention to the latter because the former is included active and exotic neutrinos, which can be very well solved through mixing matrix \cite{Hue:2017lak}. From the experimental side,the branching ratios ($Br$) of cLFV decays have upper bounds given in Ref.\cite{Patrignani:2016xqp}. 
 \bea
 Br(\mu \rightarrow e\gamma)<4.2\times 10^{-13},\crn
 Br(\tau \rightarrow e\gamma)<3.3\times 10^{-8},\crn
 Br(\tau \rightarrow \mu\gamma)<4.4\times 10^{-8}.\label{lalb-limmit}
 \eea
 
 After the discovery of the Higgs boson in 2012 \cite{Aad:2012tfa,Chatrchyan:2012ufa}, lepton-flavor-violating decays of the standard- model- like Higgs boson (LFVHDs) are getting more attention in the models  beyond the standard model (BSM). The regions of the parameter space predicted from BSM for the large  signal of LFVHDs are limited directly from both the experimental data and theory of cLFV \cite{Chatrchyan:2012ufa,Herrero-Garcia:2016uab,Blankenburg:2012ex}. Branching ratios of LFVHDs, such as ($h_1^0\rightarrow \mu\tau,\tau e$),  are stringently limited by the CMS Collaboration using data collected at a center-of-mass energy of $13\mathrm{TeV}$, as given $Br(h_1^0\rightarrow \mu\tau,\tau e)\leq \mathcal{O}(10^{-3})$. Some published results show that $Br(h_1^0\rightarrow \mu\tau)$ can reach values of $\mathcal{O}(10 ^ {-4})$ in supersymmetric and non-supersymmetric models \cite{Zhang:2015csm,Herrero-Garcia:2017xdu}. In fact, the main contributions to $Br(h_1^0\rightarrow \mu\tau)$ come from heavy neutrinos. If those contributions are minor or destructive, the $Br(h_1^0\rightarrow \mu\tau)$ in a model is only about $\mathcal{O}(10 ^ {-9})$ \cite{Gomez:2017dhl}. With the addition of heavy neutrinos, the models have many interesting features, for example, besides creating large lepton flavor violating sources, it can also explain the masses and mixing of neutrinos through the inverse seesaw (ISS) mechanism \cite{CarcamoHernandez:2019pmy, Catano:2012kw, Hernandez:2014lpa, Dias:2012xp}. Another investigation of the heavy neutrinos contribution to $Br(h_1^0\rightarrow \mu\tau)$ is also concerned, which is the mass insertion approximation (MIA) technique as shown in Ref.\cite{Marcano:2019rmk}. In this way, the interference of the contributions from the gauge and Higgs bosons is not fully presented. This leads to $Br(h_1^0\rightarrow \mu\tau)$ is far from the current experimental sensitivity.
 
 We recall that the 3-3-1 models with multiple sources of lepton flavor violating couplings have been introduced long ago \cite{PhysRevD.22.738,Chang:2006aa}. With the $\beta$ parameter, the general 3-3-1 model is separated into different layers, highlighting the properties of the neutral Higgs through contributions to rare decays that can be confirmed  by experimental data \cite{Okada:2016whh, Hung:2019jue, Dong:2008sw,  Dias:2006ns, Diaz:2004fs, Diaz:2003dk, Fonseca:2016xsy, Buras:2012dp, Buras:2014yna}. However, LFVHDs have been mentioned only in the version with heavy neutral leptons assigned as the third
 components of lepton (anti)triplets, where active neutrino
 masses are generated by effective operators \cite{Mizukoshi:2010ky, Dias:2005yh}. Another way of investigating LFVHDs is derived from the main contributions of gauge and Higgs bosons. Unfortunately, these works can only show that  the largest
 values of $Br(h_1^0\rightarrow \mu\tau)$ is $\mathcal{O}(10 ^ {-5})$ \cite{Hue:2015fbb, Thuc:2016qva}. Furthermore, the main LFV sources can also come from the charged heavy leptons as shown in the  flipped  3-3-1 model \cite{Fonseca:2016tbn}. Because the first lepton generation is arranged in a sextet, which is different from the two remaining ones. Consequently, $Br(h_1^0\rightarrow \mu\tau)$ can reach the orders of $\mathcal{O}(10 ^ {-4}-10 ^ {-3})$ when new heavy particles are in the $\mathrm{TeV}$ scale. However, $Br(l_a\rightarrow l_b\gamma)$ are much smaller than the upper bound of the experimental limit \cite{Hong:2020qxc}.
 
 Recently, the 3-3-1 model with right-handed neutrinos (331RHN)  is added heavy neutrinos which are gauge singlets (331ISS) has shown very well results when investigating LFVHDs \cite{Boucenna:2015zwa,Hernandez:2013hea,Nguyen:2018rlb}. As a result, the above model has predicted the parameter space regions, where satisfying the experimental upper limit of the $Br(\mu \rightarrow e\gamma)$ and $Br(h_1^0\rightarrow \mu\tau)$ can reach values of  $\mathcal{O}(10 ^ {-5})$ \cite{Nguyen:2018rlb}. In contrast, the 331ISS model still has some questions to be solved, such as: in the parameter space regions satisfying the experimental limits of $Br(\mu \rightarrow e\gamma)$, are $Br(\tau \rightarrow e\gamma)$ and $Br(\tau \rightarrow \mu\gamma)$ excluded? What are the contributions of neutrinos, gauge, and Higgs bosons to the LFVHDs? How does the parameterization of neutrinos mixing matrices (both active and exotic neutrinos) affect LFVHDs? In  this work, we will solve those problems.
 
 The paper is organized as follows. In the next section, we review the model and give masses spectrum of gauge and Higgs bosons. We then show the masses spectrum of the neutrinos through the inverse seesaw mechanism in Section \ref{ISSme}. We calculate the Feynman rules and analytic formulas for cLFV and LFVHDs  in Section \ref{analytic}. Numerical results are discussed in Section.\ref{Numerical}. Conclusions are in Section \ref{conclusion}. Finally, we provide Appendix \ref{DeltaLR}, \ref{CaDV} to calculate and exclude divergence in the amplitude of LFVHDs.
 \section{\label{model} The review model}
 \subsection{Particle content}
We now consider the 331ISS  model, which is structured from the original  331RHN model as given in Ref. \cite{Chang:2006aa} and additional heavy Majorana neutrinos.  The electric charge operator corresponding to  the electroweak group  $SU(3)_L\otimes U(1)_X$ is $Q=T_3+\beta T_8+X$, where $\beta=-\frac{1}{\sqrt{3}}$ and $T_{3,8}$ are diagonal $SU(3)_L$ generators.

 To avoid chiral anomalies, the left-handed components of leptons and third generations of quark are arranged into triplets, the two remaining generations of quark are in anti-triplets of $SU(3)_L$. The right-handed components of all fermions are singlets of $SU(3)_L$. Therefore, the  ($SU(3)_C;\, SU(3)_L;\, U(1)_X$) group structure of fermions are: 
 \begin{align}
   L'_{aL}& =%
 \begin{pmatrix}
 \nu' _{a} \\ 
 l'_{a} \\ 
 \left( N' _{a}\right) ^{c} \\ 
 \end{pmatrix}%
 _{L}:(1,3,-1/3), & & 
l'_{aR}:(1,1,-1) %
 ,  \notag \\
 Q'_{\al L}& =%
 \begin{pmatrix}
 d'_{\al} \\ 
 -u'_{\al} \\ 
 D'_{\al} \\ 
 \end{pmatrix}%
 _{L}:(3,3^{\ast },0), & & 
 \begin{cases}
 d'_{\al R}:(3,1,-1/3) \\ 
 u'_{\al R}:(3,1,2/3) \\ 
 D'_{\al R}:(3,1,-1/3) \\ 
 \end{cases}%
 ,  \notag \\
 Q_{L}^{\prime 3}& =%
 \begin{pmatrix}
 u'_{3} \\ 
 d'_{3} \\ 
 U' \\ 
 \end{pmatrix}%
 _{L}:(3,3,1/3), & & 
 \begin{cases}
 u'_{3R}:(3,1,2/3) \\ 
 d'_{3R}:(3,1,-1/3) \\ 
 U'_{R}:(3,1,2/3) \\ 
 \end{cases},%
 \end{align}%

 where $U'_{L}$ and $D'_{\al L}$ for $\al=1,2$ are three up- and down-type quark components in the flavor basis, while $N_{aL}^{\prime c} \cong N'_{aR}$ are right-handed neutrinos added in the bottom of lepton triplets. \\
  The scalar sector consists of a triplet $\chi $, which provides the  masses to the new heavy fermions, and two triplets $\rho $ and $\eta $,
 which give masses to the SM fermions at the electroweak scale. These scalar fields are assigned to the following $(SU(3)C; SU(3)L; U(1)X) $  representations. 
 \begin{eqnarray}
  \eta =%
 \begin{pmatrix}
 \eta_1^0   \\ 
 \eta _{2}^{-} \\ 
  \eta_3^0%
 \end{pmatrix}%
 :(1,3,-1/3);
 \hs
 \rho =%
 \begin{pmatrix}
 \rho_{1}^{+} \\ 
 \rho_2^0  \\ 
 \rho _{3}^{+} \\ 
 \end{pmatrix}%
 :(1,3,2/3);  \hs
 \chi =%
 \begin{pmatrix}
\chi_1^0 \\ 
 \chi_{2}^{-} \\ 
\chi_3^0  \\ 
 \end{pmatrix}%
 :(1,3,-1/3).  
   \label{scalar_spectrum1}
 \end{eqnarray}
  To avoid the mixing between normal and exotic quarks which may induce  dangerous flavor neutral changing currents, a soft-broken discrete $Z_2$ symmetry was introduced in Ref.~\cite{Okada:2016whh}. Particularly, all  right-handed exotic quarks  and $\chi$ are odd under this symmetry, while the remaining are even.  The Yukawa terms generating quark masses were discussed in detail in Ref.~\cite{Okada:2016whh}, but they are irrelevant to our work, therefore will not be mentioned  here. 
 
 There are two triplets ($\eta, \chi$) that have the same quantum numbers and different from a remain. Neutral components of scalar triplets are shown relevantly with real and pseudo scalars as:
 \bea 
 \eta_1^0 &=&\frac{1}{\sqrt{2}}(u+R_{1 }+ iI_{1 }); \hs \eta_3^0= \frac{1}{\sqrt{2}}(R_{1 }^\prime+ iI_{1 }^\prime) \crn
 \rho_2^0 &=&\frac{1}{\sqrt{2}}(v+R_{2 }+ iI_{2}); \hs \chi_1^0 = \frac{1}{\sqrt{2}}(R_{3 }^\prime+ iI_{3 }^\prime) ; \hs
 \chi_3^0 = \frac{1}{\sqrt{2}}(\omega +R_{3 }+ iI_{3 })  \label{scalar_spectrum2}.
 \eea
 The expectation vacuum values (vev)    $\langle \eta^0_3\rangle=\langle \chi^0_1\rangle=0 $ are assumed based on the discussion of the total lepton numbers  introduced in Refs.~\cite{Chang:2006aa,Tully:2000kk}. The 331RN model exits two global symmetries, namely $L$ and $\mathcal{L}$ are the normal and new lepton numbers, respectively. They are related to each other by $L=\frac{4}{\sqrt{3}}T_8+\mathcal{L}$. Accordingly, the normal lepton number $L$ of $\eta^0_1$, $\rho^0_2$, and $\chi^0_3$ are zeros. In contrast, $\eta^0_3$ and $\chi^0_1$ are bilepton with $L=2$. Hence,  their vacuum expectation values (VEVs) are zeros to avoid a large violation the total lepton number $L$.
 
 The electroweak symmetry breaking (EWSB) mechanism follows
 \begin{equation*}
 {SU(3)_{L}\otimes U(1)_{X}\xrightarrow{\langle \chi \rangle}}{%
 	SU(2)_{L}\otimes U(1)_{Y}}{\xrightarrow{\langle \eta \rangle,\langle \rho
 		\rangle}}{U(1)_{Q}},
 \end{equation*}
 where VEVs satisfy the hierarchy ${\omega \gg u,v}$  as done
 in Refs. \cite{Dong:2008sw,Dong:2010gk}.   In order to generate heavy neutrino masses at tree-level and arise  mixing angles, we use the ISS mechanism. Thus, the 331RHN model is extended, where
three right-handed neutrinos which are singlets of  $SU(3)_L$, $F'_{aR} \sim (1,1, 0), a = 1; 2; 3$ are added \cite{Nguyen:2018rlb}.  Requiring $\mathcal{L}$ must be softly broken, one adds a Lagrangian term, which is relevant to $F_a$ fields. The general  Lagrangian Yukawa relates to leptons and heavy neutrinos are given follows:
 \bea 
 -\mathcal{L}^{\mathrm{Y}}_{LF} =h^e_{ab}\overline{L'_{aL}}\rho l'_{bR}-
 h^{\nu}_{ab} \epsilon^{ijk} \overline{(L'_{aL})_i}(L'_{bL})^c_j\rho^*_k+Y_{ab}\overline{L'_{aL}}\,\chi F'_{bR}+\frac{1}{2} (\mu_{F})_{ab}\overline{(F'_{aR})^c}F'_{bR}
 +\mathrm{H.c.}\label{Yul}
 \eea 
 Two of the first term in Eq.(\ref{Yul})  generate masses for original charged leptons and neutrinos. The next term describes mixing between $N'_a$ and $F'_a$, and the fourth term generates masses for Majorana neutrinos $F'_a$.\\
To use the simple Higgs spectra where the mass and state of  SM-like Higgs boson will be determined exactly at the tree level, we will choose  the Higgs potential discussed in Refs. \cite{Hue:2015mna, Hue:2015fbb}, namely
  \bea
 \mathcal{V} &=& \mu_1^2 \left( \rho^\dagger \rho + \eta^\dagger \eta \right) + \mu_2^2 \chi^\dagger \chi + \lambda_1 \left( \rho^\dagger \rho + \eta^\dagger \eta \right)^2 + \lambda_2 \left( \chi^\dagger \chi \right)^2 + \lambda_{3}  \left( \rho^\dagger \rho + \eta^\dagger \eta \right)  \left( \chi^\dagger \chi \right)
 \crn
 &&- \sqrt 2 f \left(\varepsilon_{ijk} \eta^i \rho^j \chi^k + \mathrm{H.c.}  \right),
 \label{potential}
 \eea
 where $\lambda_1,\, \lambda_2,\, \lambda_3$ are the  Higgs self-coupling constants, $f$ is a mass parameter and is imposed as real. The Higgs potential~\eqref{potential} obeys the  custodial symmetry  after the first breaking step \cite{Hue:2015mna}, with the general conditions shown in Ref.~\cite{Pomarol:1993mu}. The interesting feature of this potential is that  the $\rho$ parameter is guaranteed to be close unit, therefore  it satisfies the respective current experimental constraint.
 
Following Ref.~\cite{Hue:2015mna}, we summarize here a more detailed explanation how  this custodial symmetry works on the 3-3-1 after the first symmetry breaking step $SU(3)_L\times U(1)_X\to SU(2)_L\times U(1)_Y$.  The $U(1)$ charge $Y$ is identified as~\cite{Diaz:2003dk, Buras:2012dp},
\begin{equation*}
\frac{Y}{2}= -\frac{T^8}{\sqrt{3}} +X, 	
\end{equation*}
where $T_8=1/(2\sqrt{3}) \mathrm{diag}(1,1,-2)$ for a $SU(3)_L$ triplet such as $\rho$, $\eta$ and $\chi$.  The $SU(3)_L$ Higgs triplets after this symmetry breaking step will become the following $SU(2)_L$ doublets,
$$ \eta \to \eta'= \begin{pmatrix}
\eta^0_1\\
\eta^-_2	
\end{pmatrix}\sim (2, -1),\; \rho \to \rho'=\begin{pmatrix}
\rho^+_1\\
\rho^0_2	
\end{pmatrix} \sim(2, 1), \; \chi \to \chi'= \begin{pmatrix}
\chi^0_1\\
\chi^-_2	
\end{pmatrix}\sim(2, -1),
 $$
 where the $Y$ charges of the Higgs doublets are  determined  as $Y/2=- 1/6 +X$, where   $-T^8/\sqrt{3}\to -I_2/6$. As a consequence, the $Y$ charges  of $\rho'$, $\eta'$, and $\chi'$  are 
 $$  \frac{Y_{\rho'}}{2}=-\frac{1}{6}+\frac{2}{3}= \frac{1}{2}, \; \frac{Y_{\chi'}}{2}=\frac{Y_{\eta'}}{2}= -\frac{1}{6} -\frac{1}{3}= -\frac{1}{2}.$$  
 Now at the SM scale, the two Higgs doulets $\rho'$ and $\eta'$ play roles as those appearing in the two Higgs doublet model,  that can be applied to impose a custodial on  the 3-3-1 model under consideration, using the same derivation discussed in Ref.~\cite{Hue:2015mna}, where the bi-doublet is defined as $\Phi=(\eta', \rho')/\sqrt{2}$.  Starting from the  general  Higgs potential given in Ref.~\cite{Okada:2016whh}, which contains two more terms than that considered in Ref.~\cite{Hue:2015mna}, we require that: after the first breaking step, the general Higgs potential must be identified with the following Higgs potential respecting the $SU(2)_L\times SU(2)_R$ symmetry~\cite{Pomarol:1993mu}:
 \begin{align}
 \label{eq_Vsu2LR}
 V^\mathrm{SM}_C &=m_1^2 \mathrm{Tr}\left( \Phi^\dagger \Phi\right) + \left[ m_2^2 \det(\Phi) +\mathrm{h.c.}\right] +\lambda \left[ \mathrm{Tr}\left( \Phi^\dagger \Phi\right) \right]^2 +\lambda_4 \det\left( \Phi^\dagger \Phi\right) \crn&+ \left[ \lambda_5 \left( \det \Phi\right)^2 +\mathrm{h.c.} \right]  +\left[ \lambda_6 \det(\Phi) \mathrm{Tr}\left(\Phi^\dagger \Phi\right) +\mathrm{h.c.} \right]. 
 \end{align} 
Repeating the same intermediate steps done in Ref.~\cite{Hue:2015mna}, we  obtain  the simple form of the Higgs potential given in Eq.~\eqref{potential}, which respects the custodial symmetry. The mass eigenstate and mass of the SM-like Higgs boson  are  easily to be determined analytically  with this Higgs potential. 
 
 \subsection{Gauge and Higgs bosons} 
  In the model under consideration, we denote $g$ and $g'$ as the coupling constants of the electroweak symmetry ($SU(3)_L\otimes U(1)_X$). Then, the relation between coupling constants and sine of the Weinberg angle following:
  \be  g=e\, s_W, \hs \frac{g'}{g}= \frac{3\sqrt{2}s_W}{\sqrt{3-4s^2_W}}.  \ee
    
  Gauge bosons in this model get masses through the covariant kinetic term of the Higgs bosons,
 \bea \mathcal{L}^{H}=\sum_{H=\eta,\rho,\chi} \left(D_{\mu}H\right)^{\dagger}\left(D_{\mu}H\right).\nn\eea
  
 The model comprises two pairs of singly charged gauge bosons, denoted as  $W^{\pm}$  and $Y^{\pm}$, defined as
 \bea W^{\pm}_{\mu}&=&\frac{W^1_{\mu}\mp i W^2_{\mu}}{\sqrt{2}},\hs m_W^2=\frac{g^2}{4}\left(u^2+v^2\right),\crn
 Y^{\pm}_{\mu}&=&\frac{W^6_{\mu}\pm i W^7_{\mu}}{\sqrt{2}},\hs m_Y^2=\frac{g^2}{4}\left(w^2+v^2\right). \label{singlyG}\eea
 The bosons $W^{\pm}$ as the first line in Eq.(\ref{singlyG}) are identified with the SM ones, leading to $u^2+v^2\equiv v_0^2=(246\, \mathrm{GeV})^2$.  In the remainder of the text, we will consider in detail the simple case $u=v=v_0/\sqrt{2}=\sqrt{2}m_W/g$ given in Refs. \cite{Hue:2015mna, Hue:2015fbb, Thuc:2016qva}. Under these imposing conditions, we get $h^0_1$ mixed with the three original states.\\
   From the potential Higgs was given in Eq.(\ref{potential}), one can find the masses and the mass eigenstates of Higgs bosons. There are two pair of charged Higgs $H^\pm_{1,2}$ and Goldstone bosons of $W^\pm$ and $Y^\pm$, which are denoted as $G_W^\pm$ and $G_Y^\pm$ , respectively. The relations between the original and physics states of the charged
   Higgs bosons are :
 \begin{eqnarray}
 \left( \begin{array}{c}
 \rho_1^\pm \\
 \eta_2^\pm
 \end{array} \right)= \dfrac{1}{\sqrt 2}\left( \begin{array}{cc}
 -1 & 1\\
 1 & 1
 \end{array} \right) \left(\begin{array}{c}
 G_W^\pm
 \\ H_1^\pm
 \end{array} \right), \hs \left( \begin{array}{c}
 \rho_3^\pm \\
 \chi_2^\pm
 \end{array} \right) = \left( \begin{array}{cc}
 - s_\alpha & c_\alpha \\
 c_\alpha & s_\alpha
 \end{array} \right) \left(\begin{array}{c}
 G_Y^\pm
 \\ H_2^\pm
 \end{array} \right),
 \label{EchargedH}
 \end{eqnarray}
 with masses
 \bea 
 m_{H_1^{\pm}}^2=2f\om,\hs m_{H_2^{\pm}}^2=f\om(t^2_\alpha +1), \hs  m_{G_{W}^\pm}= m_{G_{Y}^\pm}=0,\crn
 \text{and} \hs c_\alpha \equiv \cos\alpha, \hs s_\alpha \equiv \sin\alpha, \hs t_\alpha \equiv \tan\alpha=\frac{v}{\om}.
 \eea 
 With components of scalar fields are constructed as Eq.(\ref{scalar_spectrum2}), the model contains four physical CP-even Higgs bosons $h^0_{ 1;2;3;4}$ and a Goldstone boson of the non-Hermitian gauge boson $(X^0)$. The mixing of neutral Higgs $h_4^0$ and Goldstone of boson $X^0$  depends on $\gamma$ angle, ($t_\gamma \equiv \tan\gamma =\frac{u}{\omega}$). Three CP-even Higgs  $h^0_{ 1;2;3}$ mutually mix and relate to their original components as:
 \bea
 \left(\begin{array}{c} R_1\\ R_2\\ R_3 \end{array}\right) =
 \left(
 \begin{array}{ccc}
 	-\fr{c_\beta}{\sqrt{2}} & \fr{s_\beta}{\sqrt{2}} & -\fr{1}{\sqrt{2}} \\
 	-\fr{c_\beta}{\sqrt{2}} & \fr{s_\beta}{\sqrt{2}} & \fr{1}{\sqrt{2}} \\
  	s_\beta & c_\beta & 0 \\
 \end{array}
 \right)\left(\begin{array}{c} h^0_1\\ h^0_2\\ h^0_3 \end{array}\right),
 \label{mixing_CP_even_Higgs}
 \eea
 where $s_\beta = \sin\beta$ and $c_\beta = \cos\beta$, and they are defined by
 \bea
 s_\beta &=& \fr{(4 \lambda_1-m^2_{h^0_1}/v^2)t_{\alpha} }{A},\,
 c_\beta= \fr{\sqrt{2}\left(\lambda_{3}-\frac{f}{w}\right)}{A},\crn  A&=&\sqrt{ \left(4 \lambda_1 -m^2_{h^0_1}/v^2\right)^2 t_{\alpha}^2+2\left(\lambda_{3}-\frac{f}{w}\right)^2}.
 \label{angle_beta}
 \eea
There is one neutral CP-even Higgs boson $h^0_1$ with a mass proportional to the electroweak  scale and is identified with SM-like Higgs boson.

 \be m^2_{h^0_1}=\frac{w^2}{2}\left[4\lambda_1 t_{\alpha}^2 + 2\lambda_2 +\frac{ft_{\alpha}^2}{w}-\sqrt{8 t^2_{\alpha}\left(\frac{f}{w}-\lambda_{3}\right)^2+\left(2\lambda_2 +\frac{ft_{\alpha}^2}{w}-4 \lambda_1 t^2_{\alpha}\right)^2} \right]. \ee
 The remaining two neutral Higgs ($h_2^0,\,h_3^0$) in Eq.(\ref{mixing_CP_even_Higgs}) have masses on the electroweak symmetry breaking scale ($ m_{h_{2,3}^0}\sim \omega$), which are outside the range of LFVHDs so they are not given here.
\section{\label{ISSme} Neutrinos masses and ISS mechanism} 
 We now consider the Yukawa Lagrangian in Eq.(\ref{Yul}). Charged leptons masses $(m_a)$ are generated from the first term and in order to avoid LFV processes at tree level, we can assume $h^e_{ab}=\sqrt{2}\delta_{ab}m_a/v$. Thus, masses of originally charged leptons are $m_a=h^e_av/\sqrt{2}$.\\
The second term in Eq. (\ref{Yul}) is expanded by:
 \begin{align}
h^{\nu}_{ab} \epsilon^{ijk} \overline{(L'_{aL})_i}(L'_{bL})^c_j\rho^*_k
= 2 h^{\nu}_{ab} \left[-\overline{l'_{aL}} (\nu'_{bL})^c\rho^-_3+ \overline{l'_{aL}} (N'_{bL})^{c}\rho^-_1-\overline{\nu'_{aL}} (N'_{bL})^c\rho_2^{0*} \right] .\label{lastterm}
\end{align}
From the last term of Eq.(\ref{lastterm}), using  antisymmetric properties of $h^\nu_{ab}$  matrix and equality $\overline{N'_{aL}} (\nu'_{bL})^c=\overline{\nu'_{bL}} (N'_{aL})^c$, we can contribute a Dirac neutrino mass term $-\mathcal{L}_{\rm{mass}}^{\nu}=\overline{\nu'_L}\,m_D\, N'_R+\rm{H.c.}$, with basic, $\nu'_L\equiv(\nu'_{1L},\nu'_{2L},\nu'_{3L})^T$, $N'_R\equiv( (N'_{1L})^c, (N'_{2L})^c,(N'_{3L})^c)^T$ and $m_D$ has form $(m_D)_{ab}\equiv \sqrt{2}\, h^{\nu}_{ab}v$.\\
The third term in Eq.(\ref{Yul}) generates mass for heavy neutrinos, this consequence comes from the large value of Yukawa coupling $Y_{ab}$. To describe mixing $N_a$ and $F_a$, $(M_R)_{ab}=Y_{ab}\frac{\om}{\sqrt{2}}$ is introduced. The last term in Eq.(\ref{Yul})  violates both $L$ and $\mathcal{L}$, and hence $\mu_F$ can be assumed to
be small, in the scale of ISS models.\\
 In the basis $n'_L=(\nu'_L,N'_{L},(F'_{R})^c)^T$ and $(n'_L)^c=((\nu'_L)^c,(N'_{L})^c,F'_{R})^T$, Eq.(\ref{Yul}) derives mass matrix following. 
 
 \bea -{L}^{\nu}_{\mathrm{mass}}=\frac{1}{2}\overline{n'_L}M^{\nu}(n'_L)^c +\mathrm{H.c.}, \,\mathrm{ where }\quad M^{\nu}=\begin{pmatrix}
 	0	& m_D &0 \\
 	m^T_D	&0  & M_R^T \\
 	0& M_R& \mu_F
 \end{pmatrix}.  \label{Lnu1}\eea
In the normal seesaw form, $M^\nu$ can be written:
 \be  M^{\nu}=\begin{pmatrix}
 	0& M_D \\
 	M_D^T& M_N
 \end{pmatrix}, \; \mathrm{where} \, M_D \equiv(m_D,\, 0),\;  \mathrm{and} \; M_N=\begin{pmatrix}
 	0& M_R^T \\
 	M_R& \mu_F
 \end{pmatrix}. \label{Mnuss}\ee
To obtain masses eigenvalue and physics states of neutrinos, one can do diagonal $M^\nu$ by $9\times9$ matrix $U^\nu$:
\bea U^{\nu T}M^{\nu}U^{\nu}=\hat{M}^{\nu}=\mathrm{diag}(m_{n_1},m_{n_2},..., m_{n_{9}})=\mathrm{diag}(\hat{m}_{\nu}, \hat{M}_N). \label{diaMnu} \eea
The relations between the flavor and mass eigenstates are:

\bea n'_L=U^{\nu*} n_L, \hs \mathrm{and} \; (n'_L)^c=U^{\nu}  (n_L)^c, \label{Nutrans}
\eea
\be P_Ln'_i=n'_{iL} =U^{\nu*}_{ij}n_{jL},\; \mathrm{and}\; P_Rn'_i=n'_{iR} =U^{\nu}_{ij}n_{jR}, \hs i,j=1,2,...,9. \label{Nutrans2}\ee
In general,  $U^{\nu}$ is written in the form   \cite{Ibarra:2010xw},
\be U^{\nu}= \Omega \left(
\begin{array}{cc}
	U & \mathbf{O} \\
	\mathbf{O} & V \\
\end{array}
\right), \hs
\label{Unuform}\ee
\be \Omega=\exp\left(
\begin{array}{cc}
	\mathbf{O} & R \\
	-R^\dagger & \mathbf{O} \\
\end{array}
\right)=
\left(
\begin{array}{cc}
	1-\frac{1}{2}RR^{\dagger} & R \\
	-R^\dagger &  1-\frac{1}{2}R^{\dagger} R\\
\end{array}
\right)+ \mathcal{O}(R^3).
\label{Ommatrix}\ee
The matrix $U=U_{\mathrm{PMNS}}$ is the Pontecorvo-Maki-Nakagawa-Sakata (PMNS) matrix \cite{Maki:1962mu, Pontecorvo:1957qd},
\bea
U_{\mathrm{PMNS}}=\left(
\begin{array}{ccc}
	c_{12}c_{13} & s_{12}c_{13} & s_{13} e^{-i\delta} \\
	-s_{12}c_{23}-c_{12}s_{23}s_{13}e^{i\delta} & c_{12}c_{23}-s_{12}s_{23}s_{13}e^{i\delta} & s_{23}c_{13} \\
	s_{12}s_{23}-c_{12}c_{23}s_{13}e^{i\delta} & -c_{12}s_{23}-s_{12}c_{23}s_{13}e^{i\delta} & c_{23}c_{13} \\
\end{array}
\right) \mathrm{diag}(1,\; e^{i\frac{\sigma_1}{2}},\;e^{i\frac{\sigma_2}{2}}),
\label{umns}\eea
and $c_{ab}\equiv\cos\theta_{ab}$, $s_{ab}\equiv\sin\theta_{ab}$. The  Dirac phase ($\delta$) and Majorana phases ($\sigma_1,\sigma_2$) are fixed as $\delta=\pi,\sigma_1=\sigma_2=0$. In the normal hierarchy scheme, the best-fit  values of the neutrino oscillation parameters which satisfied the $3\sigma$ allowed values are given as  \cite{Zyla:2020zbs}
\bea 
s^2_{12}&=&0.32,\; s^2_{23}=0.551,\; s^2_{13}=0.0216,\crn
\Delta m^2_{21}&=& 7.55\times 10^{-5}\;\mathrm{ eV^2},\hs  \Delta m^2_{32}= -2.50\times 10^{-3}\; \mathrm{eV^2},\label{nuosc}
\eea
where $ \Delta m^2_{21}=m^2_{n_2}-m^2_{n_1}$ and $\Delta m^2_{32}=m^2_{n_3}-m^2_{n_2}$.\\
 Hence, the following seesaw relations are valid  \cite{Ibarra:2010xw}:
\bea   R^* &\simeq& \left(-m_DM^{-1}, \quad  m_D(M_R)^{-1}\right), \label{Rs}\\
m_DM^{-1} m^T_D&\simeq&m_{\nu}\equiv U^*_{\mathrm{PMNS}}\hat{m}_{\nu}U^{\dagger}_{\mathrm{PMNS}}, \label{mnu}\\
V^* \hat{M}_N V^{\dagger}&\simeq& M_N+ \frac{1}{2}R^TR^* M_N+ \frac{1}{2} M_NR^{\dagger} R, \label{masafla}\eea
where
\be M\equiv M_R^T\mu_F^{-1}M_R. \label{deM}\ee
In the framework of the 331RHN model adding $F_a$ as flavor singlets, the Dirac neutrino mass matrix $m_D$ must be antisymmetric. From results in Ref.\cite{Boucenna:2015zwa}, with
the aim of giving regions of parameter space with large LFVHDs, $m_D$ can be chosen in the form to suit both the inverse and normal hierarchy cases of active neutrino masses as: 
\be m_D\equiv \varrho \begin{pmatrix}
	0&1  &x_{13}  \\
	-1& 0 &x_{23}  \\
	-x_{13}& -x_{23} &0
\end{pmatrix}. \label{mD1}\ee
Therefore, $m_D$ has only three independent parameters $x_{13},x_{23}$, and $\varrho=\sqrt{2}vh^\nu_{23}$.\\
In general, the matrix $m_{\nu}$ in Eq. (\ref{mnu}) is symmetric, $(m_{\nu})_{ij}=(m_{\nu})_{ji}$, their components are given by:
\bea 
(m_\nu)_{ij}=(m_D)_{ik}(M^{-1})_{kl}(m_D^T)_{lj},\quad (k\ne i,l\ne j).
\eea 
We can calculate in detail,
\bea 
(m_{\nu})_{ij}-(m_{\nu})_{ji}&\sim x_{13}\left[M^{-1}_{13}-M^{-1}_{31}\right]+ x_{23}\left[M^{-1}_{23}-M^{-1}_{32}\right]+(M^{-1}_{12}-M^{-1}_{21}),\quad \text{for}\quad  i\ne j \crn
(m_\nu)_{11}&=M^{-1}_{22}+x_{13}(M^{-1}_{23}+M^{-1}_{32})+x^2_{13}M^{-1}_{33},\crn
(m_\nu)_{22}&=-M^{-1}_{11}-x_{23}(M^{-1}_{13}+M^{-1}_{31})+x^2_{23}M^{-1}_{33},\crn
(m_\nu)_{33}&=x^2_{13}M^{-1}_{11}+x_{13}x_{23}(M^{-1}_{12}+M^{-1}_{21})+x^2_{23}M^{-1}_{22}.
\label{mnu_ij}
\eea   
From Eq.(\ref{mnu_ij}), we have two solutions $x_{13}$, $x_{23}$ and one equation, which express the relation of components of matrix $m_\nu$: 
\bea x_{13}&=\frac{(m_{\nu})_{23}\left[ (m_{\nu})_{13}^2-(m_{\nu})_{11}(m_{\nu})_{33}\right] +(m_{\nu})_{13}\sqrt{\left[ (m_{\nu})_{13}^2-(m_{\nu})_{11}(m_{\nu})_{33}\right] \left[ (m_{\nu})_{23}^2-(m_{\nu})_{22}(m_{\nu})_{33}\right] }}{(m_{\nu})_{13}^2(m_{\nu})_{22}-(m_{\nu})_{11}(m_{\nu})_{23}^2},\crn
x_{23}&=\frac{(m_{\nu})_{13}\left[ (m_{\nu})_{23}^2-(m_{\nu})_{22}(m_{\nu})_{33}\right] +(m_{\nu})_{23}\sqrt{\left[ (m_{\nu})_{13}^2-(m_{\nu})_{11}(m_{\nu})_{33}\right] \left[ (m_{\nu})_{23}^2-(m_{\nu})_{22}(m_{\nu})_{33}\right] }}{(m_{\nu})_{13}^2(m_{\nu})_{22}-(m_{\nu})_{11}(m_{\nu})_{23}^2},\crn
(m_{\nu})_{11}&(m_{\nu})^2_{23}+(m_{\nu})_{22}(m_{\nu})^2_{13}+ (m_{\nu})_{33}(m_{\nu})^2_{12}=
(m_{\nu})_{11}(m_{\nu})_{22}(m_{\nu})_{33}+ 2(m_{\nu})_{12}(m_{\nu})_{13}(m_{\nu})_{23}.\crn
\label{nxij}\eea
Based on experimental data of neutrinos oscillation in Eq.(\ref{nuosc}), the matrix $m_D$ is parameterized and only depends on $\varrho$.
\bea  m_D\simeq \varrho \times \begin{pmatrix}
	0	& 1 & 0.7248 \\
	-1	& 0 & 1.8338 \\
	-0.7248& -1.8338 &0
\end{pmatrix}. \label{nmD}\eea
It should be emphasized that $m_D$ has a form like Eq.(\ref{nmD}) and differ from Ref.\cite{Nguyen:2018rlb}. This is caused Eq.(\ref{mnu_ij}) can have many different solutions, but the use of Eq.(\ref{nmD}) is suited very well to investigate LFVHDs.
\section{\label{analytic} Couplings and analytic formulas} 
In this section, we will calculate amplitudes and branching ratios of the LFVHDs in terms of $M^\nu$ and physical neutrino masses. With this aim, all vertices are presented in term of physical masses and mixing parameters. From relation in Eq.(\ref{diaMnu}), one can derive equation as follow:
\bea
M^{\nu}_{ab}&=&\left(U^{\nu*}\hat{M}^{\nu}U^{\nu\dagger}\right)_{ab}=0 \rightarrow U^{\nu*}_{ak}U^{\nu*}_{bk}m_{n_k}=0.\label{conMnu}
\eea
Here $a,b=1,2,3$ and $m_{n_k}$ is mass of neutrino $n_k$, with $k$ run taken over $1,2,...,9$. It is interesting that the relation in Eq.(\ref{conMnu}) leads to represent Yukawa couplings in terms of $M^\nu$ and physical neutrino masses.
\bea
\sqrt{2}v\,h^{\nu}_{ab} &=& (m_D)_{ab}= (M^{\nu})_{a(b+3)}=(U^{\nu*}\hat{M}^{\nu}U^{\nu\dagger})_{a(b+3)}=U^{\nu*}_{ak}U^{\nu*}_{(b+3)k}m_{n_k},\crn
\frac{w}{\sqrt{2}} Y_{ab}&=&(M_R)_{ab}= (M^{\nu})_{(a+3)(b+6)}=U^{\nu*}_{(a+3)k}U^{\nu*}_{(b+6)k}m_{n_k}.
\label{rel1}\eea
We then pay attention to the relevant couplings of LFVHDs. These couplings are derived by  Lagrangian Yukawa, Lagrangian kinetics of lepton (or scalar) fields, and Higgs potential. From the first term in Eq.(\ref{Yul}), we can give couplings between leptons and Higgs boson as follow:
\bea
&-& h^e_{ab}\overline{L'_{aL}}\rho l'_{bR}+{\rm h.c.}=- \frac{g  m_{a}}{ m_W}\left[ \overline{\nu'_{aL}}l'_{aR}\rho^+_1 +\overline{l'_{aL}}l'_{aR}\rho_2^0+ \overline{N'_{aL}}l'_{aR}\rho^+_3+\rm{ h.c.}\right]\crn
&\supset&\frac{g\, m_{a}c_{\beta}}{2m_W} h^0_1\overline{l_{a}}l_{a}-\frac{g\, m_a}{\sqrt{2} m_W}\left[ \left(U^{\nu}_{ai} \overline{n_{i}}P_Rl_{a}H^+_1+ U^{\nu*}_{ai} \overline{l_{a}}P_Ln_{i}H^-_1\right)\right]\crn
&&-\frac{g\, m_{a}}{m_W}\left[c_{\alpha}\left(U^{\nu}_{(a+3)i} \overline{n_{i}}P_Rl_{a}H^+_2+ U^{\nu*}_{(a+3)i} \overline{l_{a}}P_Ln_{i}H^-_2\right)\right].
\label{eephi}\eea
The relevant couplings in the second term of the Lagrangian in Eq. (\ref{Yul}) are:
\bea
&& h^{\nu}_{ab} \epsilon^{ijk} \overline{(L'_{aL})_i}(L'_{bL})^c_j\rho^*_k+ \rm{h.c.}\crn
&=&2 h^{\nu}_{ab} \left[-\overline{l'_{aL}} (\nu'_{bL})^c\rho^-_3-\overline{\nu'_{aL}} (N'_{bL})^c\rho_2^{0*}+ \overline{l'_{aL}} (N'_{bL})^c\rho^-_1 \right]\crn
&=&\frac{gc_{\beta}}{2\,m_W}h^0_1\left[ \sum_{c=1}^3U^{\nu}_{ci}U^{\nu*}_{cj}\overline{ n_i}\left(m_{n_i}P_L+m_{n_j}P_R\right)n_j \right]\crn
&-&\frac{gc_{\alpha}}{m_W} \left[ (m_D)_{ab}U^{\nu}_{bi} H^-_2\overline{l_{a}}P_Rn_i+\rm{h.c.}\right]+\frac{g}{\sqrt{2}m_W} \left[ (m_D)_{ab}U^{\nu}_{(b+3)i} H^-_1\overline{l_{a}}P_Rn_i+\rm{h.c.}\right].
\label{psipsiphi}\eea
The couplings get contributions of $M_R$ matrix given by:
\bea&-&Y_{ab}\overline{L'_{aL}}\,\chi F'_{bR}+\rm{h.c.}\crn
&=&-\frac{\sqrt{2}}{w} (M_R)_{ab}\left[ \overline{\nu'_{aL}}\chi^0_1 +\overline{l'_{aL}}\chi_2^{-} + \overline{N'_{aL}}\chi^0_3\right] F'_{bR} +\rm{h.c.}\crn
&\supset& -\frac{g t_{\alpha}} {\sqrt{2}m_W}(M_R)_{ab}\left[s_{\beta} U^{\nu}_{(a+3)i} U^{\nu}_{(b+6)j}\overline{n_{i}}P_Rn_{j}h^0_1+ \sqrt{2}s_{\alpha}U^{\nu}_{(b+6)i} \overline{l_{a}}P_Rn_{i}H^-_2 +\rm{h.c.} \right]. \label{NRYe}\eea
Neutrinos interact with gauge bosons based on the kinetic terms of the leptons. When we only concern with the couplings of the charged gauge bosons, the results are:
\bea  \mathcal{L}^{\ell\ell V}=\overline{L'_{aL}}\gamma^{\mu}D_{\mu}L'_{aL}
&\supset&\frac{g}{\sqrt{2}} \left( \overline{l'_{aL}}\gamma^\mu  \nu'_{aL}W^{-}_{\mu} + \overline{l'_{aL}}\gamma^\mu  N'_{aL} Y^{-}_{\mu} \right)+\mathrm{h.c.}\crn
&=&\frac{g}{\sqrt{2}} \left[ U^{\nu*}_{ai} \overline{l_{a}}\gamma^\mu P_L  n_{i}W^{-}_{\mu} + U^{\nu}_{ai} \overline{ n_{i}}\gamma^\mu P_L l_{a}W^{+}_{\mu}\right.\crn
&+& \left.U^{\nu*}_{(a+3)i}\overline{l_{a}}\gamma^\mu  P_L n_{i} Y^{-}_{\mu} + U^{\nu}_{(a+3)i}\overline{n_{i} }\gamma^\mu  P_L l_{a}Y^{+}_{\mu} \right]. \label{llv1}\eea
To calculate $h_1^0\overline{ n_{i}}n_i$ coupling, we use results in Eqs.(\ref{psipsiphi}, \ref{llv1}) as given above. Furthermore, we can define the symmetry coefficient $\lambda^0_{ij}=\lambda^0_{ji}$ as Ref.\cite{Nguyen:2018rlb}, the result therefore obtained.
\be \lambda^0_{ij}=\sum_{k=1}^3\left(U^{\nu}_{ki}U^{\nu*}_{kj}m_{n_i}+U^{\nu*}_{ki}U^{\nu}_{kj}m_{n_j}\right)-\sum_{k,q=1}^3
\sqrt{2} t_{\alpha}t_{\beta}(M^*_R)_{cd} \left[U^{\nu*}_{(k+3)i} U^{\nu*}_{(q+6)j}+U^{\nu*}_{(k+3)j} U^{\nu*}_{(q+6)i}\right].\crn
\ee
This result is a coincidence with the Feynman rules given in Ref.\cite{Dreiner:2008tw}. In such way, the $h^0_1\overline{ n_{k}}n_j$ coupling can be written in the symmetric form $h^0_1\overline{ n_{k}}n_j \sim h^0_1\overline{ n_{k}}(\lambda^0_{kj}P_L+\lambda^{*0}_{jk}P_R)n_j$. For brevity, we also define the coefficients related to the interaction of charged Higgs and fermions as follows:
 \bea \lambda^{L,1}_{ak}&=& -\sum_{i=1}^3(m_D^*)_{ai}U^{\nu*}_{(i+3)k},\quad 
 \lambda^{R,1}_{ak}=m_{a}U^{\nu}_{ak},
\crn
\lambda^{L,2}_{ak}&=& \sum_{i=1}^3\left[(m_D^*)_{ai}U^{\nu*}_{ik}+ t^2_{\alpha}(M_R^*)_{ai}U^{\nu*}_{(i+6)k}\right],\quad \lambda^{R,2}_{ak}=m_{a}U^{\nu}_{(a+3)k}.
\eea
The couplings related to LFVHDs  are given in Tab.(\ref{numbers}). Especially, based on the characteristics of this model, some couplings of $h^0_1$ such as  $h_1^0H_1^\pm H_2^\mp$,\,$h_1^0Y^\pm W^\mp$,\,$h_1^0Y^\pm H_1^\mp$,\,$h_1^0W^\pm H_{1,2}^\mp$ are zeros.
  \begin{table}[ht]
	\begin{tabular}{|c|c|}
		\hline
		Vertex & Coupling \\
		\hline
		$ h^0_1 \overline {l_a}l_a$ & $\frac{igm_{a}}{2m_W}c_{\beta}$ \\
		\hline
		$ h^0_1 \overline {n_k}n_j$ & $\frac{igc_{\beta}}{2m_W}\left(\lambda^0_{kj}P_L+\lambda^{0*}_{kj}P_R\right)$ \\
			\hline
		$h^0_1 H^+_2H^-_2$ &$ i\lambda^{\pm}_{H_2}= - i w \left[  2 s_\beta s_\alpha^2 \lambda_2 +s_\beta c_\alpha^2 \lambda_{3} - \sqrt 2 \left( 2 c_\beta c_\alpha^2 \lambda_1 + c_\beta s_\alpha^2 \lambda_{3}  \right) t_\alpha - \dfrac{\sqrt 2}{\om} f c_\beta c_\alpha s_\alpha \right]$  \\
		\hline
		$ h^0_1H^+_1H^-_1$ &$i\lambda^{\pm}_{H_1}=-i v \left(-2 \sqrt 2 c_\beta \lambda_1 + \dfrac{s_\beta v_3 \lambda_{3} + s_\beta f}{v}  \right) $\\
		\hline
		$  H_2^+\overline{n_k} l_b$, 	$H_2^-\overline {l_a} n_k $ & $-\frac{igc_{\alpha}}{m_W}\left(\lambda^{L,2}_{bk}P_L+\lambda^{R,2}_{bk}P_R\right)$,  $-\frac{igc_{\alpha}}{m_W}\left(\lambda^{L,2*}_{ak}P_R+\lambda^{R,2*}_{ak}P_L\right)$\\
		\hline
		$  H_1^+\overline{n_k} l_b$, 	$H_1^-\overline {l_a} n_k $ & $-\frac{ig}{\sqrt{2}m_W}\left(\lambda^{L,1}_{bk}P_L+\lambda^{R,1}_{bk}P_R\right)$,  $-\frac{ig}{\sqrt{2} m_W}\left(\lambda^{L,1*}_{ak}P_R+\lambda^{R,1*}_{ak}P_L\right)$\\	
		\hline
		$ W_\mu^+ \overline{ n_k} l_b$,  	$   W_\mu^-\overline {l_a}n_k$ & $\frac{ig}{\sqrt{2}}U^{\nu}_{bk}\gamma^\mu P_L$, $\frac{ig}{\sqrt{2}}U^{\nu*}_{ak}\gamma^\mu P_L$\\
		\hline
		$Y_\mu^+ \overline{n_k}  l_b$, 	$Y_\mu^-  \overline {l_a} n_k $& $\frac{ig}{\sqrt{2}}U^{\nu}_{(b+3)k}\gamma^\mu P_L$, $\frac{ig}{\sqrt{2}}U^{\nu*}_{(a+3)k}\gamma^\mu P_L$\\
		\hline
		$Y_{\mu}^-H_2^+h^0_1$, $Y^+_{\mu}H_2^-h^0_1$ &$\dfrac{ig}{2 \sqrt 2} \left( c_\al c_\beta + \sqrt 2 s_\al s_\beta \right)  \left( p_{h_1^0} - p_{H_2^+} \right)^\mu$, $  \dfrac{ig}{2 \sqrt 2} \left( c_\al c_\beta + \sqrt 2 s_\al s_\beta \right)  \left( p_{H_2^-} - p_{h_1^0} \right)^\mu $  \\
		\hline
		$h^0_1 W^+_{\mu }W^-_{\nu}$ &  $ -ig m_W c_\beta\,g^{\mu\nu} $\\
		\hline
		$h^0_1 Y^+_{\mu }Y^-_{\nu}$ & $ \frac{ig m_Y}{\sqrt 2} \left( \sqrt 2 s_\beta c_\alpha - c_\beta s_\alpha \right) g^{\mu\nu}$\\
			\hline
	\end{tabular}
	\caption{Couplings related to the SM-like Higgs decay ($h^0_1\rightarrow l_al_b$) in the 331ISS model. All momenta in the Feynman rules corresponding  to these vertices are incoming.  \label{numbers}}
\end{table}

\begin{figure}[ht]
	\centering
	\begin{tabular}{cc}
		\includegraphics[width=14.0cm]{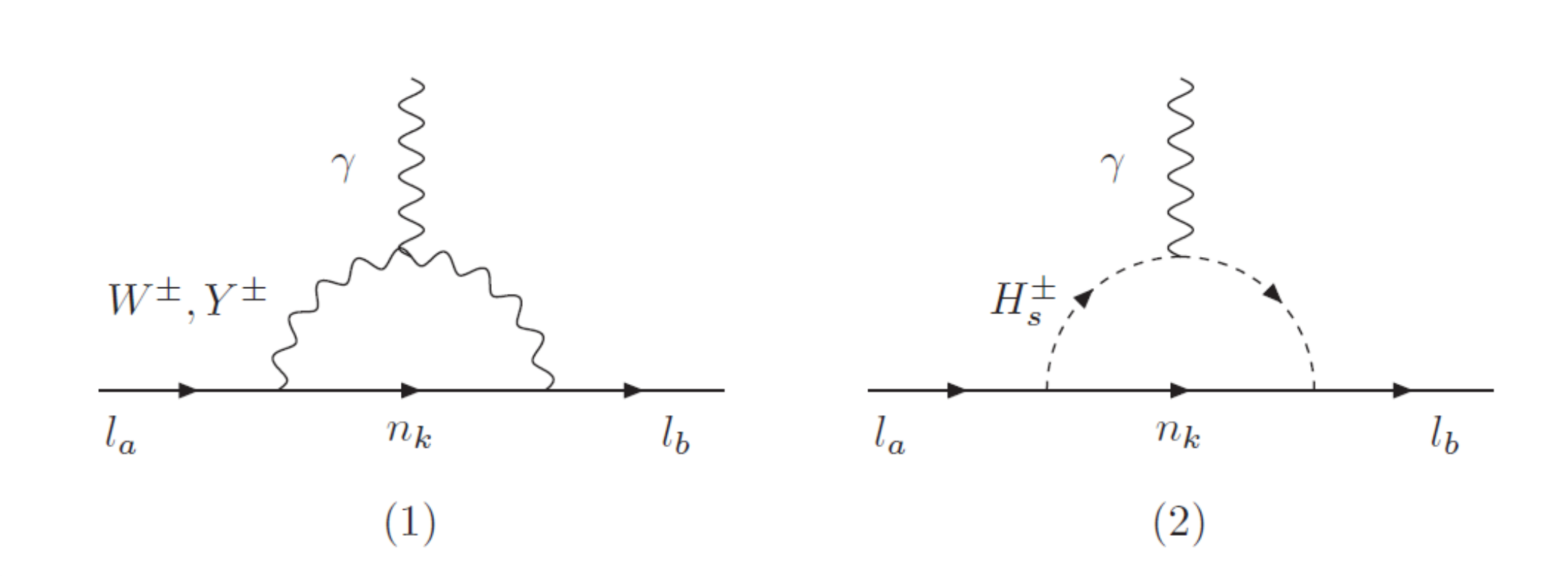} 
	\end{tabular}%
	\caption{ Feynman diagrams at one-loop order of $l_a \rightarrow l_b \gamma$ decays in the unitary gauge. In diagram ($2$), $H_s^\pm$ ($s=1,2$) is charged Higgs bosons in this model}
	\label{fig_eiejga}
\end{figure}

From Tab.(\ref{numbers}), we can show all Feynman diagrams at one-loop order  of the $l_a \rightarrow l_b \gamma$ decays in the unitary gauge as Fig.(\ref{fig_eiejga}).

 The regions of parameter space predicting large branching ratios ($Br$) for LFVHDs are affected strongly by the current
experimental bound of $\mathrm{Br}(l_a \rightarrow l_b \gamma)$, with $(a>b)$ \cite{TheMEG:2016wtm}. Therefore, we will simultaneously investigate the LFV decay of charged leptons and LFVHDs. In the limit $m_{a, b} \rightarrow 0$,  where $m_{a, b}$ are denoted for the masses of charged leptons $l_{a, b}$, respectively, we can derive the result, which is a very good approximate formula for branching ratio of cLFV as given in Ref.\cite{Hue:2017lak}
\begin{equation}\label{brmuega}
\mathrm{Br}(l_a \rightarrow l_b\gamma)\simeq \frac{12\pi^2}{G_F^2}|D_R|^2\mathrm{Br}(l_a\rightarrow l_b\overline{\nu_b}\nu_a),
\end{equation}
where $D_R=D_R^{W^\pm}+D_R^{Y^{\pm}}+D_R^{H_s^{\pm}}$ and $G_F=\frac{g^2}{4\sqrt{2}m_W^2}$ . The analytic forms are represented as:
\begin{align}
D^{W^\pm}_R&=-\frac{eg^2}{32\pi^2m_W^2} \sum_{k=1}^9U^{\nu*}_{ak}U^{\nu}_{bk}F(t_{kW}),\crn
D^{Y^\pm}_R&=  -\frac{eg^2}{32\pi^2m_Y^2} \sum_{k=1}^9U^{\nu*}_{(a+3)k}U^{\nu}_{(b+3)k}F(t_{kY}), \crn
D^{H^{\pm}_s}_R&=-\frac{eg^2f_s}{16\pi^2m_W^2} \sum_{k=1}^9\left[\frac{ \lambda^{L,s*}_{ak}  \lambda^{L,s}_{bk}}{m^2_{H^{\pm}_s}}\times\frac{1-6t_{ks} +3 t^2_{ks} +2t^3_{ks} -6t^2_{ks} \ln(t_{k_s})}{12 (t_{ks}-1)^4} \right.\crn
&+\left. \frac{m_{n_k} \lambda^{L,s*}_{ak} \lambda'^{R,s}_{bk}}{m^2_{H^{\pm}_s}}\times \frac{-1 +t_{ks}^2 -2t_{ks} \ln(t_{ks})}{2(t_{ks}-1)^3} \right].\label{DRlalbga}
\end{align}
Some quantities are defined as below. Especially, the $ F(x)$ function is derived from the characteristics of PV functions in this model.
\begin{align}
t_{kW} &\equiv \frac{m^2_{n_k}}{m_W^2},\; t_{kY}\equiv \frac{m^2_{n_k}}{m_Y^2}, \; t_{ks}\equiv \frac{m^2_{n_k}}{m^2_{H^{\pm}_s}},\crn
f_1&\equiv c^2_{\alpha},\; f_2\equiv \frac{1}{2},\;   \lambda'^{R,1}_{bk}\equiv U^{\nu}_{bk}, \; \lambda'^{R,2}_{bk}\equiv U^{\nu}_{(b+3)k},\crn
F(x)&\equiv -\frac{10-43x+78x^2-49x^3 +4x^4 +18x^3\ln(x)}{12(x-1)^4}.
\end{align}
For different charge lepton decays, we use experimental data $\mathrm{Br}(\mu\rightarrow e\overline{\nu_e}\nu_\mu)=100\%, \mathrm{Br}(\tau\rightarrow e\overline{\nu_e}\nu_\tau)=17.82\%, \mathrm{Br}(\tau\rightarrow \mu\overline{\nu_\mu}\nu_\tau)=17.39\% $ as given in Ref.\cite{Patrignani:2016xqp}.

 Apart from the cLFV decays $l_b\rightarrow \l_a \gamma$, which Br$(\mu\rightarrow e\gamma) <4.2\times 10^{-13}$  is  normally considered as the most stringent  experimental constraint, there are some other types of cLFV decays    such as the $l\rightarrow l'l'' l'''$ and the $\mu-e$ conversion in nuclei, for example see a review in ref.~\cite{Lindner:2016bgg} including  discussions for the 331RN model with a Higgs sextet. In the model under consideration, the leading contributions to these processes are one-loop level, and all relevant LFV couplings also contribute to the decay amplitudes $l_b\rightarrow l_a \gamma$. Therefore,  theoretical constraint can be established naively as  Br$(\mu\rightarrow 3e),\; \mathrm{CR}(\mu^-\; \mathrm{Ti}\rightarrow e^- \mathrm{Ti}) \le  \mathcal{O}(10^{-2}) \mathrm{Br}(\mu\rightarrow e\gamma)$~\cite{Lindner:2016bgg}. These two processes have the most stringent  constraints from experiments, Br$(\mu \rightarrow 3e)<10^{-12}$~\cite{Bellgardt:1987du} and CR$(\mu^-\; \mathrm{Ti}\rightarrow e^- \mathrm{Ti})<6.1\times 10^{-13}$, which are weaker than that from Br$(\mu \rightarrow e\gamma)$. Therefore, it is enough to  look for the regions of the parameter space  satisfying the recent constraints of Br$(l_b\rightarrow l_a \gamma)$. In the future, the experimental constraints from the decays $l\rightarrow l'l'' l'''$ and the $\mu-e$ conversion in nuclei may be more stringent than the Br$( \mu \rightarrow e \gamma)$.  This is a very interesting topic  which should be investigated in detail in a separated work.

\begin{figure}[ht]
	\centering
	\begin{tabular}{cc}
		\includegraphics[width=15.0cm]{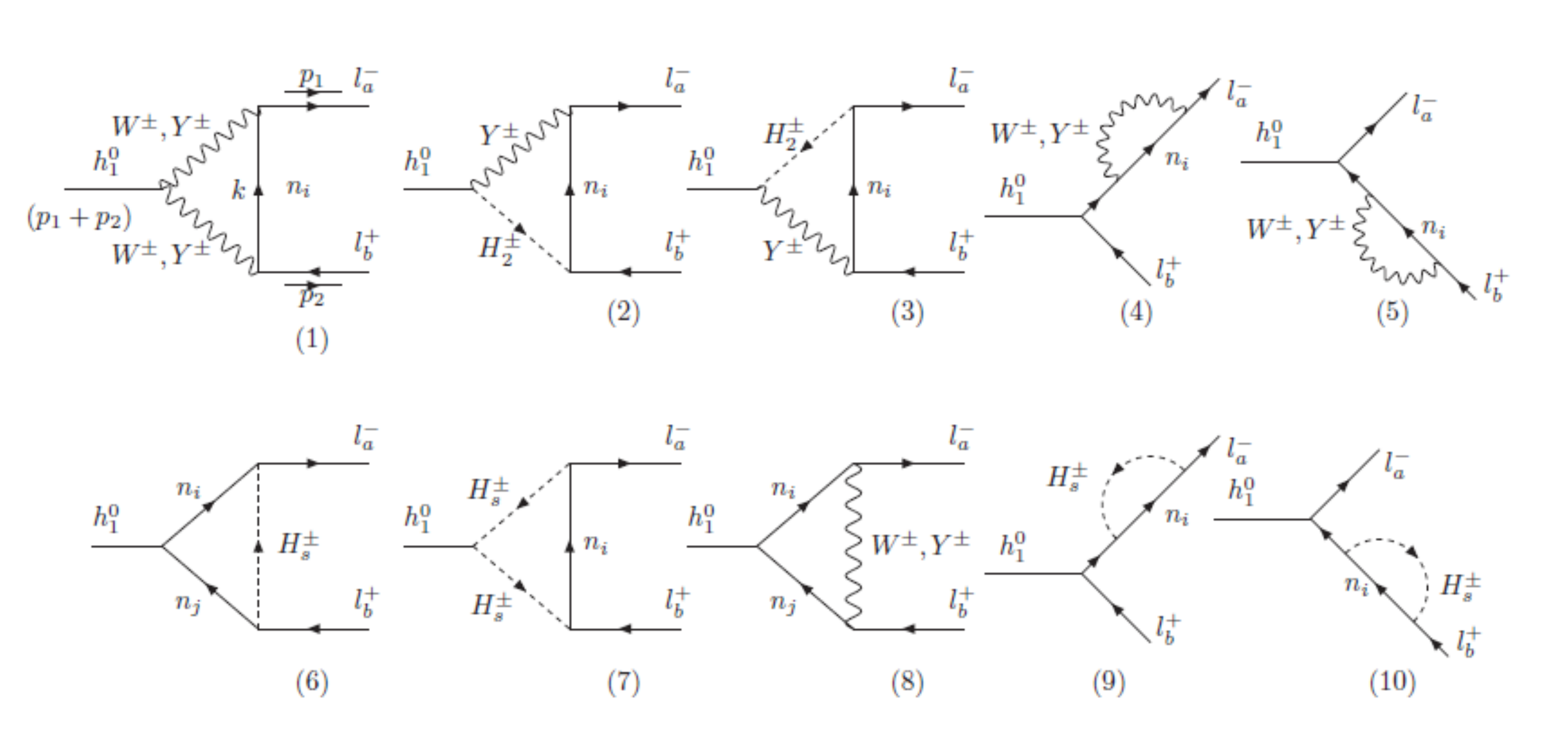} 
	\end{tabular}%
	\caption{ Feynman diagrams at one-loop order of $h_1^0 \rightarrow l_al_b$ decays in unitary gauge.}
	\label{fig_hmt}
\end{figure}

To investigate  the LFVHDs of the SM-like Higgs boson  $h^0_1\rightarrow l_a^{\pm}l_b^{\mp}$, we use scalar factors $\Delta_{(ab)L}$ and $\Delta_{(ab)R}$, which was first obtained in Ref.\cite{Nguyen:2018rlb}. Therefore, the effective Lagrangian of  these decays is
$ \mathcal{L}_{\mathrm{LFVH}}^\mathrm{eff}= h^0_1 \left(\Delta_{(ab)L} \overline{l_a}P_L l_b +\Delta_{(ab)R} \overline{l_a}P_R l_b\right) + \mathrm{h.c.}$
Based on the couplings in Tab.\ref{numbers}, we obtain the one-loop Feynman diagrams contributing to the LFVHDs amplitude
in the unitary gauge are shown in Fig.\ref{fig_hmt}. The scalar factors $\Delta_{(ab)L,R}$  arise from the loop contributions. Here, we only pay attention to corrections at one-loop order.\\
The partial width of $h^0_1\rightarrow l_a^{\pm}l_b^{\mp}$ is:
\be
\Gamma (h_1^0\rightarrow l_al_b)\equiv\Gamma (h^0_1\rightarrow l_a^{+} l_b^{-})+\Gamma (h_1^0\rightarrow l_a^{-} l_b^{+})
=  \fr{ m_{h^0_1} }{8\pi }\left(\vert \Delta_{(ab)L}\vert^2+\vert \Delta_{(ab)R}\vert^2\right), \label{LFVwidth}
\ee
with conditions $p^2_{1,2}=m^2_{a,b}$,\, $(p_1+p_2)^2=m^2_{h^0_1}$ and $m^2_{h^0_1}\gg m^2_{a,b}$.\, We obtain branching ratio is $Br(h_1^0\rightarrow l_al_b)=\Gamma (h_1^0\rightarrow l_al_b)/\Gamma^\mathrm{total}_{h^0_1}$, where $\Gamma^\mathrm{total}_{h^0_1}\simeq 4.1\times 10^{-3}\mathrm{GeV}$\cite{Patrignani:2016xqp,Denner:2011mq}. All Feynman diagrams at one-loop order in unitary gauge contributing to $h_1^0 \rightarrow l_al_b$ decay are given follow.

Factors contribute to the partial width of $h^0_1\rightarrow l_a^{\pm}l_b^{\mp}$ are
\be \Delta_{(ab)L,R} =\sum_{k=1,4,5,8} \Delta^{(k)W}_{(ab)L,R} + \sum_{k=1,2,3,4,5,8} \Delta^{(k)Y}_{(ab)L,R}+ \sum_{k=6,7,9,10} \Delta^{(k)H_s}_{(ab)L,R},  \label{deLR}\ee
where the analytic forms of $\Delta^{(k)W,Y,H_s}_{(ab)L,R}$ are calculated using the unitary
gauge and shown in the App.\ref{DeltaLR}. The amplitudes of each diagram (denoted by $k$) in Fig.\ref{fig_hmt} are represented analytically by PV (Passarino -Veltman) functions. In which, only $C_i$ functions, $i=0,1,2$,  are finite functions, the rest are diverging. However, the divergence cancellation of the total amplitude in Eq.(\ref{deLR}) is proved analytically  by techniques similar to Ref.\cite{Kuipers:2012rf, Nguyen:2018rlb} and presented as App.\ref{CaDV}. Here, we show the term groups for which the divergence has been eliminated. 
\bea
\Delta_{1,L,R}&=&\Delta^{(1)W}_{(ab),L,R}+\Delta^{(8)W}_{(ab),L,R}+\Delta^{(6)H_1}_{(ab),L,R}+\Delta^{(9+10)H_1}_{(ab),L,R},\crn
\Delta_{2,L,R}&=&\Delta^{(1)Y}_{(ab),L,R}+\Delta^{(2)Y}_{(ab),L,R}+\Delta^{(3)Y}_{(ab),L,R}+\Delta^{(8)Y}_{(ab),L,R}+\Delta^{(6)H_2}_{(ab),L,R}+\Delta^{(9+10)H_2}_{(ab),L,R},\crn
\Delta_{3,L,R}&=&\Delta^{(7)H_1}_{(ab),L,R}+\Delta^{(7)H_2}_{(ab),L,R}+\Delta^{(4+5)W}_{(ab),L,R}+\Delta^{(4+5)Y}_{(ab),L,R}.\label{del_i}
\eea
Based on the finite terms $\Delta_{1,L,R}, \Delta_{2,L,R}, \Delta_{3,L,R}$ ($\Delta_i,\,i=\overline{1,3}$ - for short), we can investigate the change in total amplitude of $h_1^0 \rightarrow l_al_b$ with the masses of the heavy neutrinos and other  parameters of the model.
\section{\label{Numerical} Numerical results of cLFV and LFVHD} 
\subsection{\label{Numerical1}Setup parameters}  
We use the well-known experimental parameters \cite{Zyla:2020zbs,Patrignani:2016xqp}: 
the charged lepton masses $m_e=5\times 10^{-4}\,\mathrm{GeV}$,\,  $m_\mu=0.105\,\mathrm{GeV}$,\, $m_\tau=1.776\,\mathrm{GeV}$,\, the SM-like Higgs mass $m_{h^0_1}=125.1\,\mathrm{GeV}$,\,  the mass of the W boson $m_W=80.385\,\mathrm{GeV}$  and the gauge coupling of the $SU(2)_L$ symmetry $g \simeq 0.651$.\\
To numerically investigate the $l_a \rightarrow l_b\gamma$ and the LFVHDs, we choose the free
parameters  are: mass of charged gauge boson $m_Y$, Higgs self-coupling constants $\lambda_1,\, \lambda_3$, mass of charged Higgs $m_{H^\pm_1}$. Therefore, the dependent parameters are given as follows:
\bea 
v&=\frac{\sqrt{2}m_W}{g}, \, s_\alpha=\frac{m_W}{\sqrt{2}m_Y},\,\omega=\frac{2m_Y}{gc_\alpha},\crn
f&=\frac{gc_\alpha m^2_{H^\pm_1}}{4m_Y},\, m^2_{H^\pm_2}=\frac{m^2_{H^\pm_1}}{2}\left(t^2_\alpha +1 \right) ,\crn
\lambda_2&=\frac{t^2_\alpha }{2}\left(\frac{m^2_{h^0_1}}{v}-\frac{m^2_{H^\pm_1}}{2\omega^2} \right) +\frac{\left( \lambda_3-\frac{m^2_{H^\pm_1}}{2\omega^2}\right) ^2}{4\lambda_1-\frac{m^2_{h^0_1}}{v^2}} .\label{deppara}
\eea 
The charged gauge
boson mass $m_Y$ is related to the lower constraint of
neutral gauge boson $Z'$ in 3-3-1 models, which have also been mentioned in Refs.\cite{Buras:2013dea, Salazar:2015gxa}. To satisfy those constraints, we choose the default value $ m_Y = 4.5\, \mathrm{TeV}
$. The values of  the Higgs self-couplings must satisfy theoretical conditions of
unitarity and the Higgs potential must be bounded from
below,   which also guarantee
that all couplings of the SM-like Higgs boson approach the
SM limit when $v\ll \omega$. For the above reasons, the Higgs self-couplings are fixed as $\lambda_1=1,\, \lambda_3=-1$. Based on recent data of neutral meson
mixing $B_0 - \overline{B}_ 0$  \cite{Okada:2016whh}, we can choose the lower bound of $m_{H^\pm_1}\geq 500\mathrm{GeV}$. This is also consistent with Ref.\cite{Nguyen:2018rlb}. Characteristic for the scale of the matrix $m_D$ is the parameter $\varrho$ as shown in Eq.(\ref{nmD}), considered in the range of the perturbative limit, $\varrho =\sqrt{2}vh^\nu_{23}\leq 617 \,\mathrm{GeV}$. In the calculations below, we fix the values for $\varrho$ to be: $100$, $200$, $400$, $500$ and $600\, \mathrm{GeV}$. To represent masses of heavy neutrinos ($F_a$), we parameterize the matrix $M_R$ in the form of a diagonal. In particular, the hierarchy of a diagonal matrix $M_R$ can yield large results for the LFVHDs.
\subsection{\label{Numerical2} Numerical results of cLFV}   
\begin{figure}[ht]
	\centering
	\begin{tabular}{cc}
		\includegraphics[width=7.8cm]{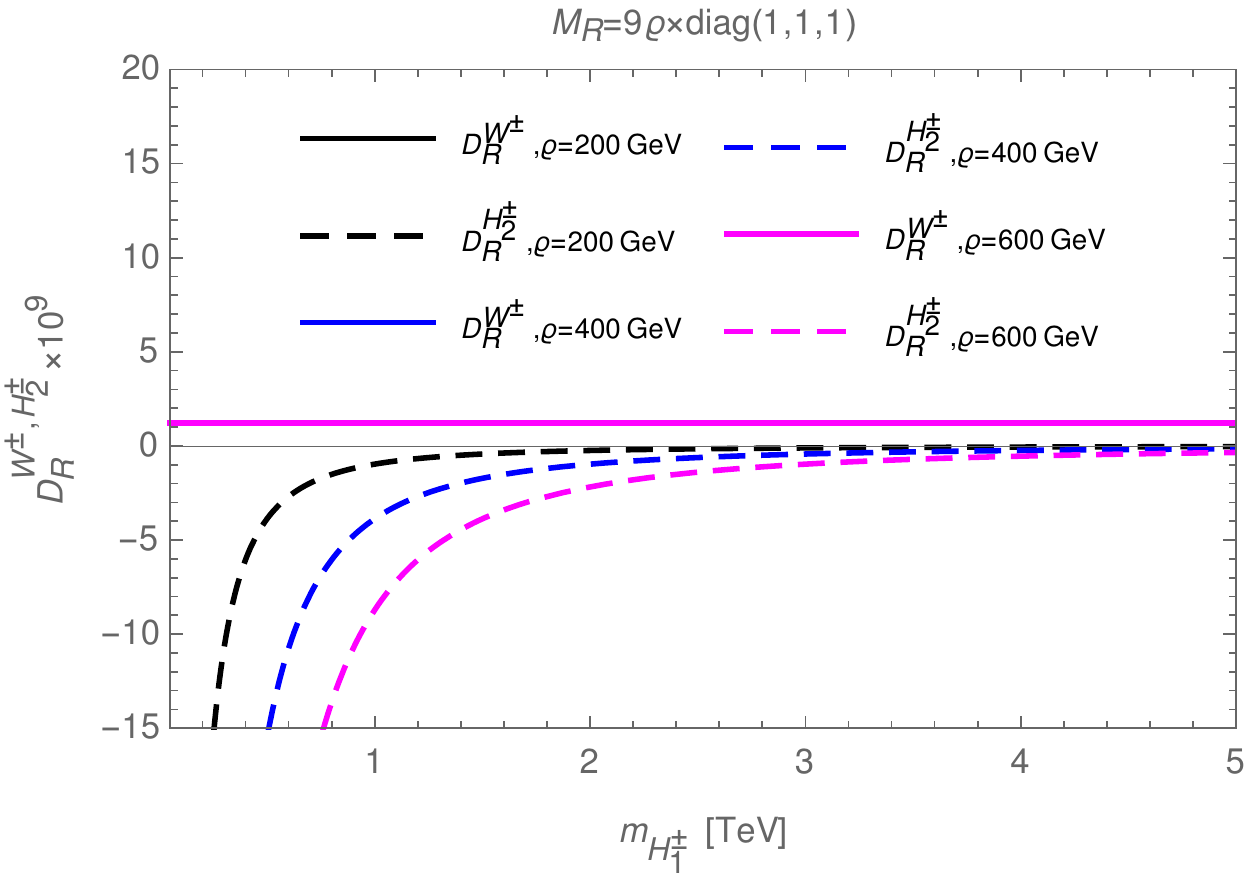} &
		\includegraphics[width=7.8cm]{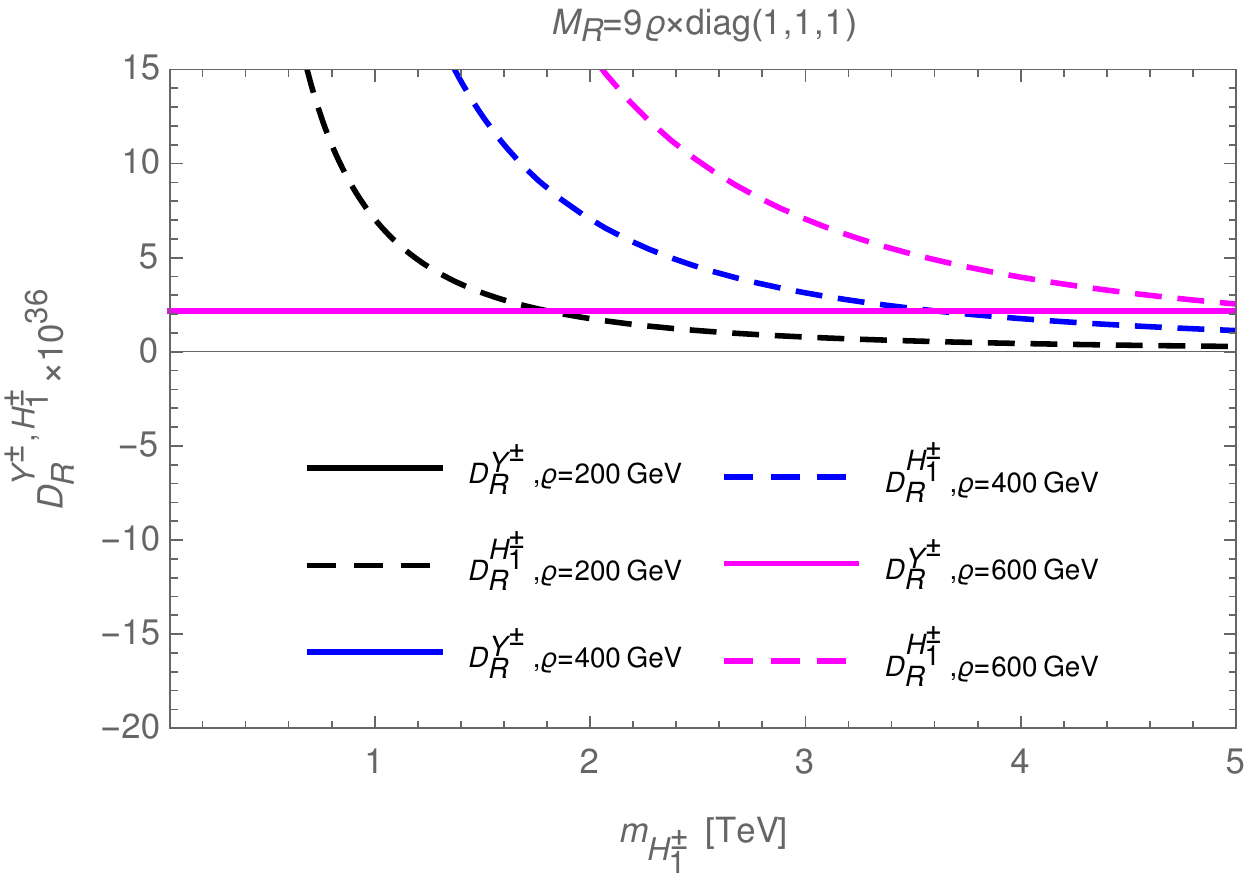} \\
	\end{tabular}%
	\caption { Contributions of $D_R^{W^\pm}$, $D_R^{H_2^\pm}$ (left) and $D_R^{Y^\pm}$, $D_R^{H_1^\pm}$ (right) to $Br(\mu \rightarrow e\gamma)$ as function of $m_{H_1^\pm}$ with fixed $\varrho = 200, 400, 600\,\mathrm{GeV}$.
		\label{fig_DR}}
\end{figure}

In this section, we numerically investigate of  $l_a \rightarrow l_b\gamma$ decays with $a> b$  use a diagonal and non-hierarchical $M_R$  matrix. That means $M_R\sim diag(1,1,1)$. We overhaul regions mentioned in Ref.\cite{Nguyen:2018rlb}, where $\varrho$ was chosen to be from a hundred of GeV to $600\mathrm{GeV}$,
and $M_R$ was in form $M_R=k\varrho diag(1,1,1)$, with $k$ is small. As a result, it is shown the existence of  the narrow regions of parameter space where can satisfy the experimental bound on $Br(\mu \rightarrow e\gamma)<4.2\times 10^{-13}$ and change fastly with the change of $m_{H_1^\pm}$. \\
To indicate the origin, we numerically investigate the  contributions to $Br(\mu \rightarrow e\gamma)$ in Eq.(\ref{DRlalbga}). We choose $M_R=9\varrho \; diag(1,1,1)$, $\varrho$ is fixed $200, 400, 600\,\mathrm{GeV}$ and $m_{H_1^\pm}$ is in range $(0.1,\, 5)\mathrm{TeV}$. The contributions of gauge and Higgs bosons defend on $m_{H_1^\pm}$ as shown in Fig.\ref{fig_DR}.

In the parameter space under consideration, $D_R^{W^\pm}$ and $D_R^{H_2^\pm}$ have the same order of size $10 ^{- 9}$ (left), while $D_R^{Y^\pm}$ and $D_R^{H_1^\pm}$ are of size $10 ^{- 36}$ (right). Therefore, the contributions of $D_R^{W^\pm}$ and $D_R^{H_2^\pm}$ to $Br(\mu \rightarrow e\gamma)$ are dominant parts. A salient result is that, while $D_R^{W^\pm}$ is always positive and almost unchanged for fixed values of $\varrho$, $D_R^{H_1^\pm}$ is negative and decreases as $\varrho$ increases. It is the reason that the contributions of gauge and Higgs bosons are destructive, creating the narrow regions of parameter space where satisfy $Br(\mu \rightarrow e\gamma)<4.2\times 10^{-13}$.

This is a very interesting property of this model, the main contributions at one loop order  to $Br(\mu \rightarrow e\gamma)$ are made up of charged bosons ($H^\pm_s,\,W^\pm,\,Y^\pm$) and neutrinos. These contributions are opposite in sign, leading to mutual reduction to produce values of $Br(\mu \rightarrow e\gamma)$ that satisfy the upper limit of the experiment. This consequence does not occur when only neutral bosons and exotic charged leptons are contributed as shown in Ref.\cite{Hong:2020qxc}. Because, the main contributions in that case do not create interference then in the region of parameter space where $Br(\mu \rightarrow e\gamma)$ is close to upper bound, $Br(\tau \rightarrow \mu \gamma)$ and $Br(\tau \rightarrow e \gamma)$ are much smaller than the current experimental limits.

\begin{figure}[ht]
	\centering
	\begin{tabular}{ccc}
		\includegraphics[width=5.0cm]{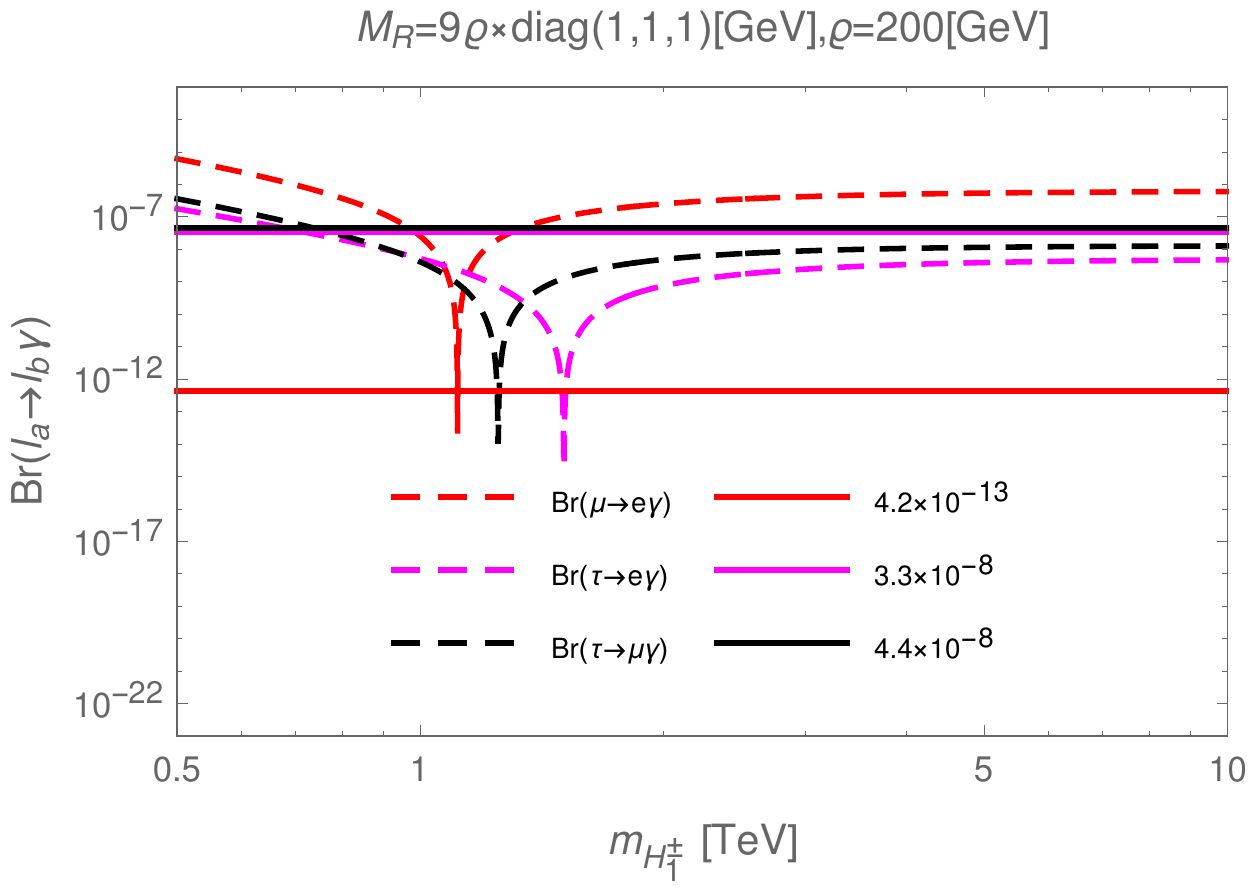} &\includegraphics[width=5.0cm]{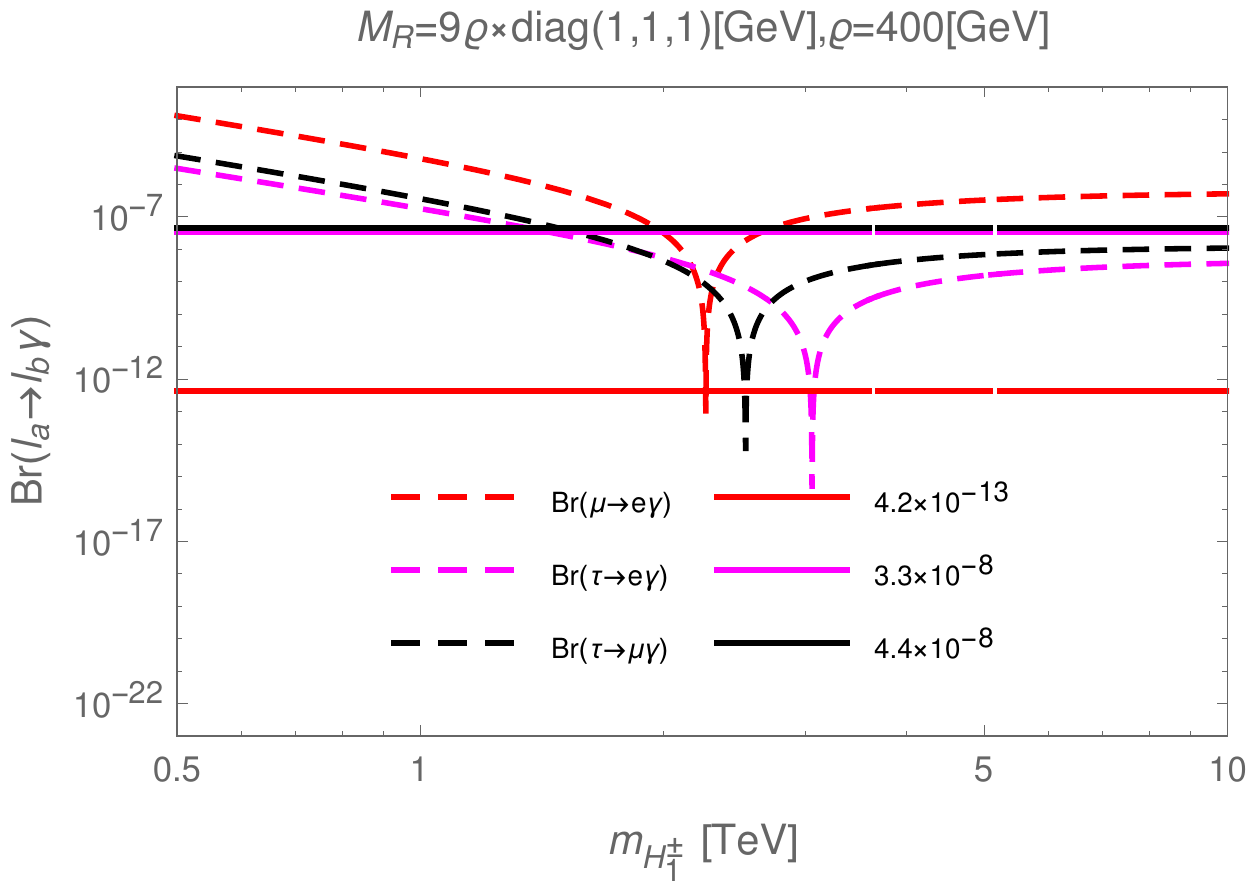} &\includegraphics[width=5.0cm]{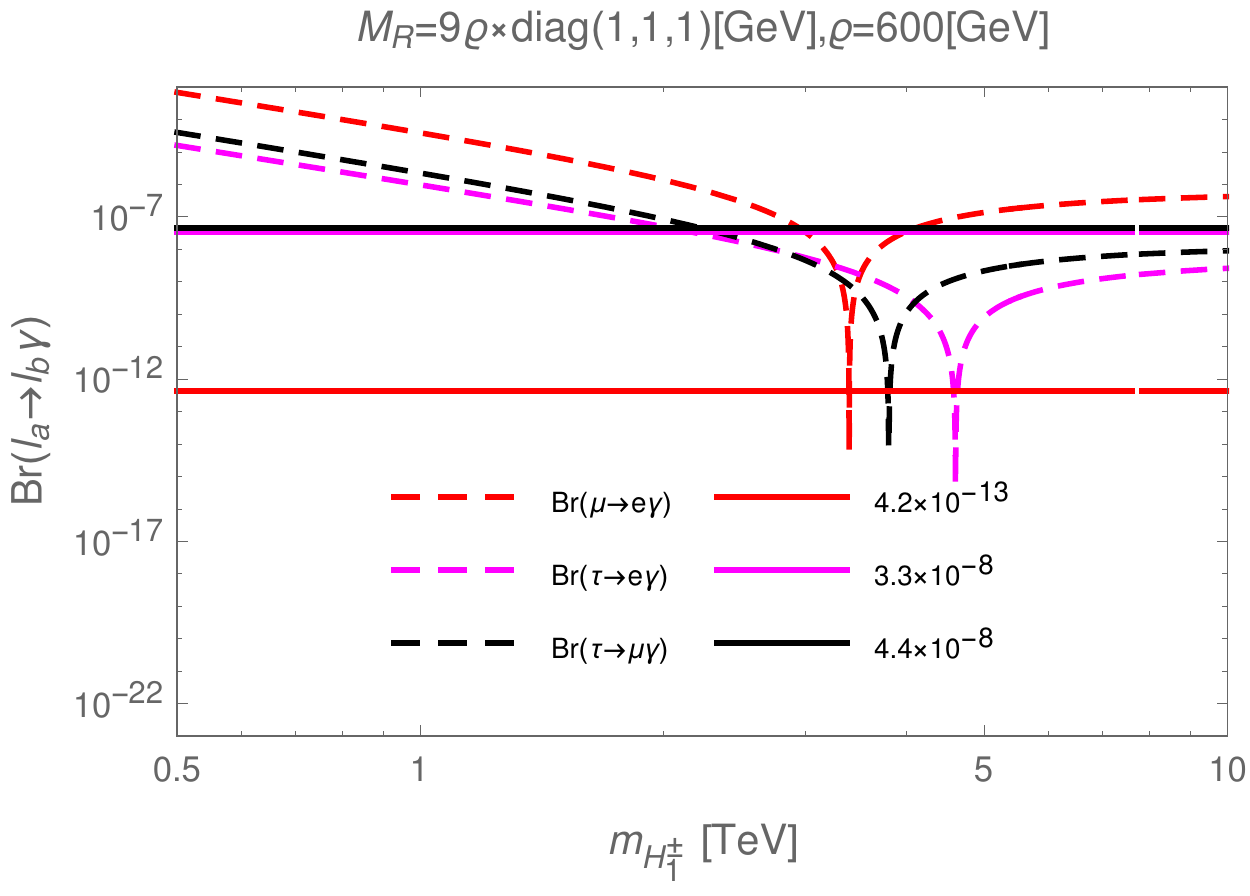} \\
    	\includegraphics[width=5.0cm]{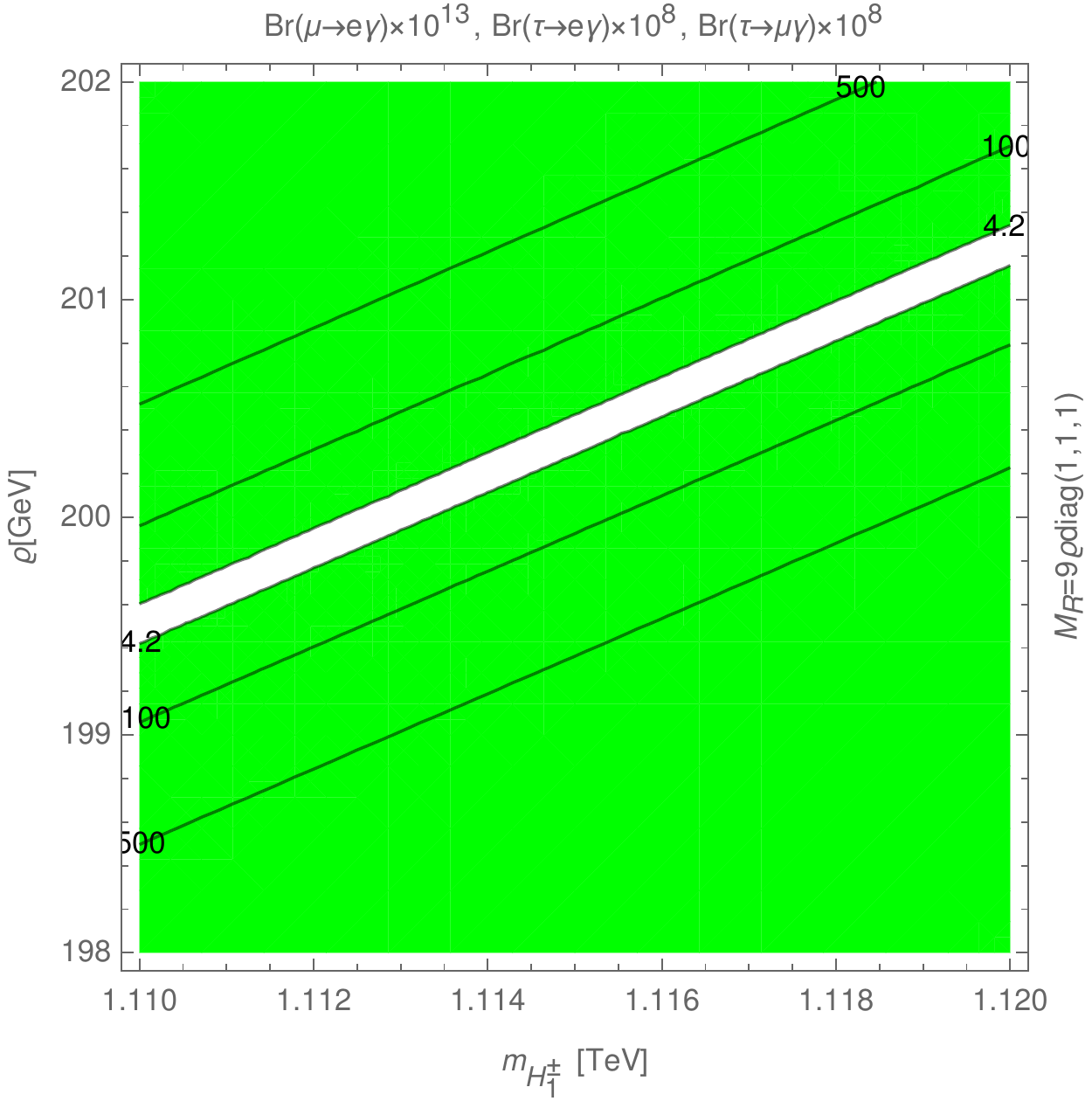} & \includegraphics[width=5.0cm]{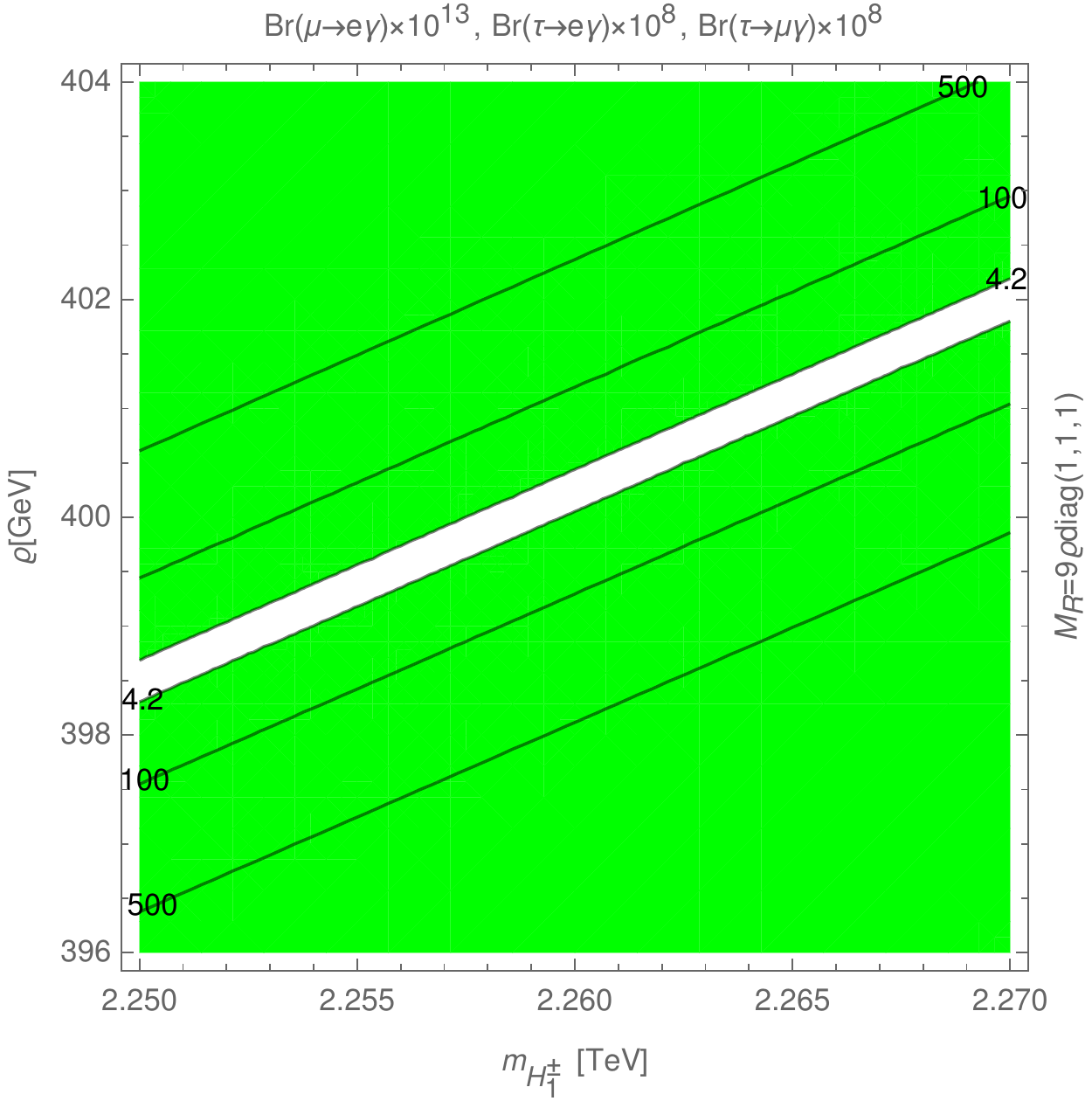} 
         & \includegraphics[width=5.0cm]{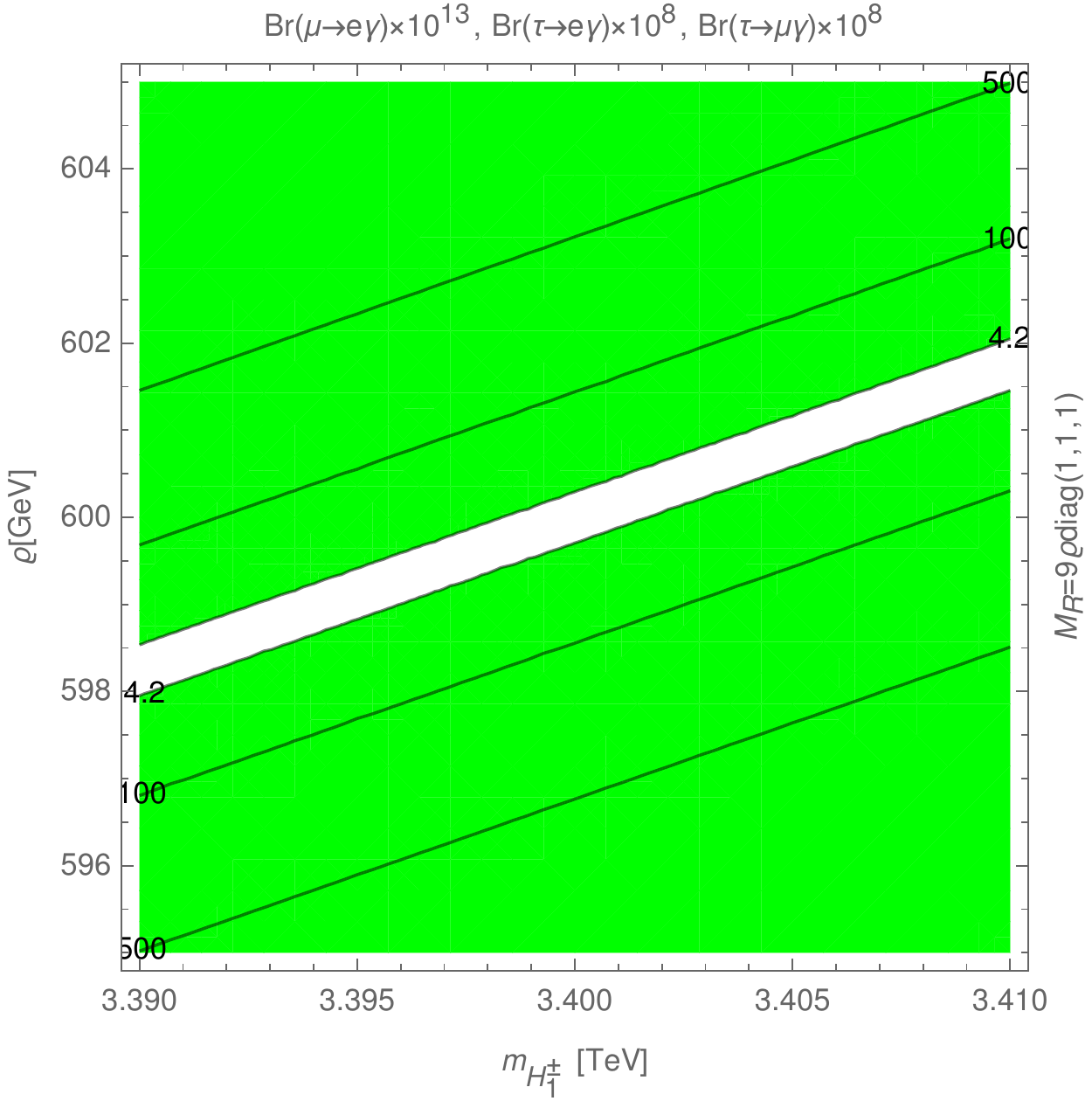} \\
  \end{tabular}%
	\caption {Plots of $Br(l_a\rightarrow l_b\gamma)$ depend on $m_{H_1^\pm}$ (first row) and contour plots of $Br(l_a\rightarrow l_b\gamma)$  as functions of $\varrho$ and $m_{H_1^\pm}$ (second row).
	\label{fig_eiej}}.
\end{figure}

In a similar way, we can investigate the contributions of gauge and Higgs bosons to $Br(\tau \rightarrow e\gamma)$ and $Br(\tau \rightarrow \mu\gamma)$ according to change of $m_{H_1^\pm}$ and find out the regions of parameter space which comply with current experimental limits $Br(\tau \rightarrow e\gamma)<3.3\times10^{-8}$ and $Br(\tau \rightarrow \mu\gamma)<4.4\times10^{-8}$ \cite{Patrignani:2016xqp}. However, the parameter space is only really meaningful when all the experimental limits are satisfied.
For the above reasons, we will examine $Br(\tau \rightarrow e\gamma)$ and $Br(\tau \rightarrow \mu\gamma)$ in narrow space regions allowed to satisfy the experimental limits of $Br(\mu \rightarrow e\gamma)$. We choose $k=9$ and fix $\varrho=200,400,600\mathrm{GeV}$, the range of $m_{H_1^\pm}$ is from $500\mathrm{GeV}$ to $10\mathrm{TeV}$, $Br(l_a \rightarrow l_b\gamma)$ defend on $m_{H_1^\pm}$ are shown in Fig.\ref{fig_eiej}.

We illustrate how $Br(l_a\rightarrow l_b\gamma)$ change with $m_{H_1^\pm}$, in the case $k = 9$ and the fixed values of $\varrho = 200,400,600\mathrm{GeV}$, corresponding to the plots in the first row of Fig.\ref{fig_eiej}.  Here, we obtain narrow parameter spaces that satisfy the experimental limits of the $Br(\mu\rightarrow e\gamma)$, respectively, shown in the second row. The expected space (colorless) is between the two curves $4.2$, the remaining is ruled out of the experimental limits (green). It is easy to see that in the narrow regions of space where $Br(\mu\rightarrow e\gamma)\leq 4.2\times 10^{-13}$, although the contributions at one loop order of bosons and neutrinos to $Br(\mu\rightarrow e\gamma)$ are mutually destructive, but  they enhance and make $Br(\tau\rightarrow \mu \gamma,(e\gamma))$ in the size of $10^{-10},(10^{-9})$. It should be emphasized that, for other fixed values of $\varrho$ within the limits of the perturbation theory, we can also investigate in the same way.

In each allowed narrow space, where the $Br(\mu\rightarrow e\gamma)$ is within the experimental limits, $Br(\tau\rightarrow e\gamma)$ and $Br(\tau\rightarrow \mu\gamma)$ also satisfy the upper bound limits of the experiment. In particular, the values of $Br(\tau\rightarrow e\gamma)$  can reach as high as $10^{-9}$ and  $Br(\tau\rightarrow \mu\gamma)$ is about $10^{-10}$, close to the accuracy found in today's large accelerators. These results are shown in Tab.\ref{Tablalb}.\\ 
  \begin{table}[h]
	\begin{tabular}{|c|c|c|c|}
		\hline
		$\varrho[\mathrm{GeV}]$&Values of $m_{H_1^\pm}[\mathrm{TeV}]$ satisfy&  Values of $Br(\tau \rightarrow e\gamma)\times 10^{9}$&Values of  $Br(\tau \rightarrow \mu\gamma)\times 10^{10}$\\&$Br(\mu \rightarrow e\gamma)<4.2\times 10^{-13}$&&\\
		\hline
		$ 200$ &$1.112\rightarrow 1.114 $&$2.269\rightarrow 2.304$& $8.569\rightarrow 8.872$ \\
		\hline
		$ 400$ &$2.256 \rightarrow 2.261$&$2.158\rightarrow  2.198$& $7.813\rightarrow  8.161$ \\
		\hline
		$ 600$ &$3.396 \rightarrow 3.405 $&$2.124\rightarrow  2.172 $& $7.579 \rightarrow 7.986$ \\
		\hline
	\end{tabular}
	\caption{The ranges of $Br(\tau\rightarrow e\gamma)$ and $Br(\tau\rightarrow \mu\gamma)$ in narrow space regions where the experimental limits of $Br(\mu\rightarrow e\gamma)$ are satisfied. \label{Tablalb}}
\end{table}

The $l_a \rightarrow l_b \gamma$ processes have also been studied previously in the context of the 3-3-1 models such as Ref.\cite{Cabarcas:2013jba}. According to the result, the parameter space areas satisfy the experimental limits (comply with Refs.\cite{Hayasaka:2010np,Lees:2010ez}) of the $l_a \rightarrow l_b \gamma$ are given. However, it has two restrictions: {\it i}) the limit of $Br(\mu \rightarrow e\gamma)$ is not tight ($2.4\times 10^{-12}$), {\it ii}) has not shown the regions of the parameter space suitable for all $l_a \rightarrow l_b \gamma$ decays. These restrictions have been overcome as shown in Tab.\ref{Tablalb}. This is a very interesting result given in the framework of this model and a suggestion for the verification of physical effects in the model from current experimental data.
\subsection{\label{Numerical3} Numerical results of LFVHD}   
We consider the narrow spatial regions where the $Br(l_a\rightarrow l_b\gamma)$ approach the experimental upper limit, this may be predicting large LFVHD. Therefore, we will investigate the contributions of $\Delta_{i,R,L},i=\overline{1,3}$ in Eq.(\ref{del_i}) to $Br(h^0_1 \rightarrow l_al_b)$ and then, we continue to examine $Br(h^0_1 \rightarrow l_al_b)$ in the  narrow spaces mentioned above. This is done both in the case of hierarchical and non-hierarchical $M_R$.

Without loss of generality when studying $Br(h^0_1 \rightarrow l_al_b)$, we will choose $Br(h^0_1 \rightarrow \mu \tau)$. $M_R$ matrix is chosen non-hierarchically in the form $M_R=9\varrho \times diag(1,1,1)$ and hierarchically in the form $M_R=9\varrho \times diag(1,2,3)$ and $M_R=9\varrho \times diag(3,2,1)$.

\begin{figure}[ht]
	\centering
	\begin{tabular}{cc}
		\includegraphics[width=7.1cm]{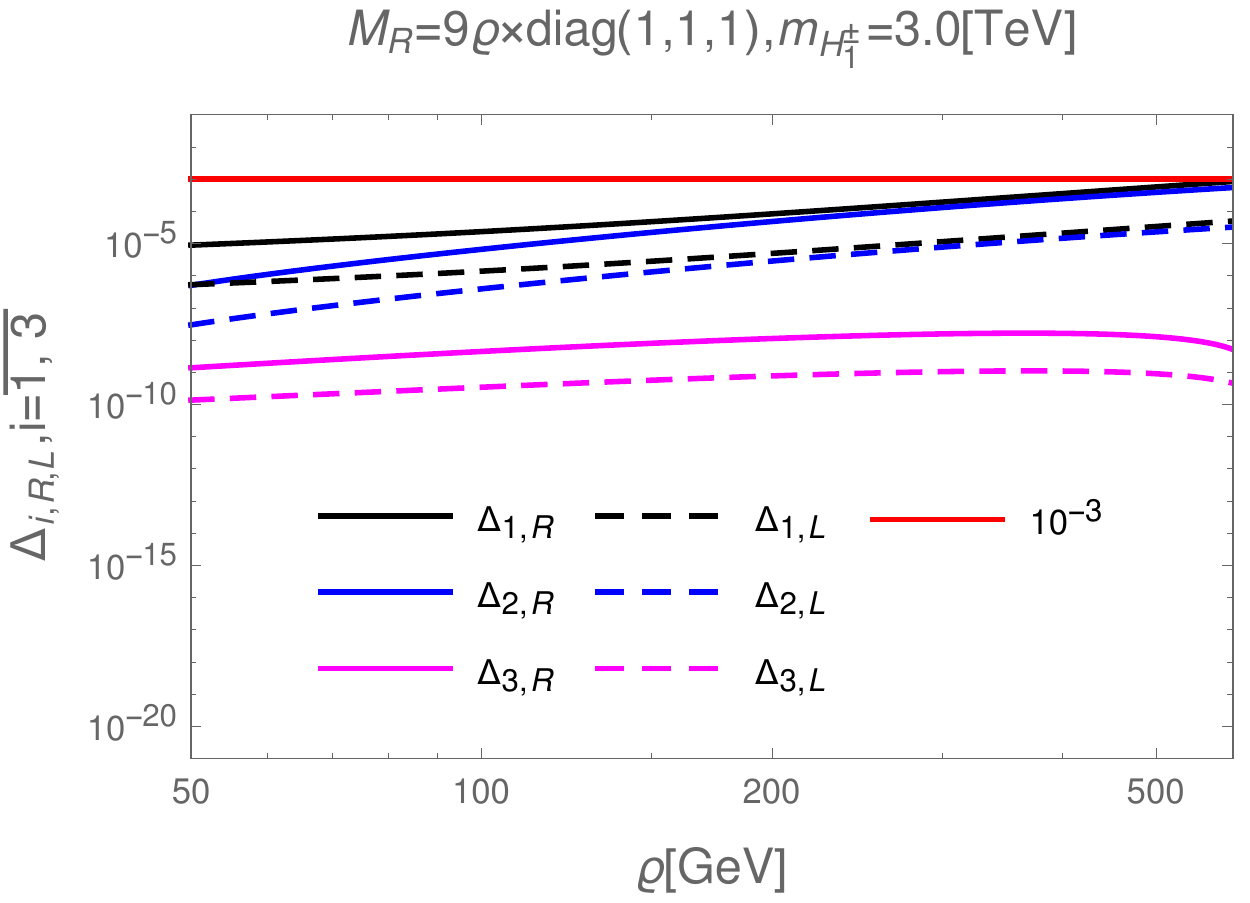} &
		\includegraphics[width=7.0cm]{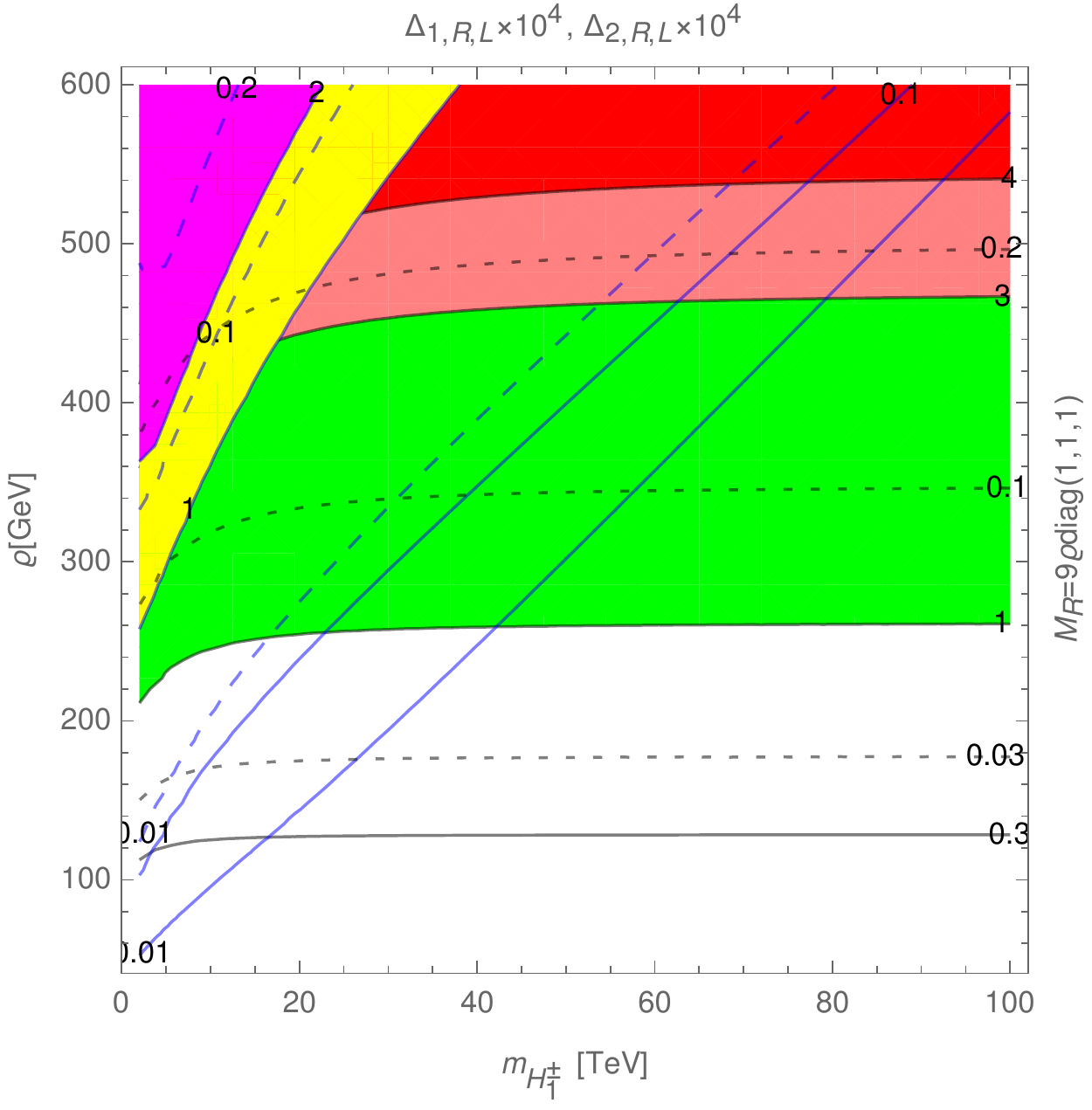} \\
	\end{tabular}%
	\caption{ Plots of $\Delta_{i,R,L},i=\overline{1,3}$  as functions of $\varrho$ with fixed value $m_{H_1^\pm}=3.0 \mathrm{TeV}$ (left) and contour plots of $\Delta_{1,2,R,L}$ as functions of $m_{H_1^\pm}$ and $\varrho$ (right). In the right panel, the black, blue, dashed-black and dashed-blue curves present constant values of $\Delta_{1,R},\Delta_{2,R},\Delta_{1,L},\Delta_{2,L}$, respectively. All plots are investigated in case $M_R=9\varrho \times diag(1,1,1)$.}
	\label{fig_con1}
\end{figure}

In case $M_R=9\varrho \times diag(1,1,1)$, the contributions of $\Delta_{i,R,L},i=\overline{1,3}$ to $Br(h^0_1 \rightarrow \mu \tau)$ are given in Fig.\ref{fig_con1}. With the parameter domain of this model selected in \ref{Numerical1}, for each fixed value of $m_{H_1^\pm}$, $\Delta_{i,R,L},i=\overline{1,3}$ increase with $\varrho$ and contribution of $\Delta_3$ was very small compared to ones of $\Delta_{1,2}$. For illustration, we choose an arbitrary value of $m_{H_1^\pm}$ ($m_{H_1^\pm}=3.0 \mathrm{TeV}$) to examine contributions of $\Delta_{i},i=\overline{1,3}$  to $Br(h^0_1 \rightarrow \mu \tau)$ with $\varrho$ is chosen in range ($50,600$)$\mathrm{GeV}$. The results are shown in the left panel of Fig.\ref{fig_con1}. Thus, we can ignore the contribution of $\Delta_{3}$ to $Br(h^0_1 \rightarrow \mu \tau)$.

In the right panel of Fig.\ref{fig_con1}, we choose $m_{H_1^\pm}$ in range ($0.5,100$) $\mathrm{TeV}$ and area of $\varrho$ is from $50 \mathrm{GeV}$ to $600 \mathrm{GeV}$. The green, pink, red present the value ranges of $1< \Delta_{1,R}\times 10^{4} < 3,3< \Delta_{1,R}\times 10^{4} < 4,\Delta_{1,R} \times 10^{4}> 4$, respectively. The yellow, magenta illustrate areas of $\Delta_{2,R} >1\times 10^{-4}$. We can find that the magenta region may gives the largest $Br(h^0_1\rightarrow \mu\tau)$, with $\Delta_{(\mu\tau)R}\sim 6\times 10^{-4}$ and $\Delta_{(\mu\tau)L}\sim 0.4\times 10^{-4}$.

\begin{figure}[ht]
 	\centering
 	\begin{tabular}{cc}
 		\includegraphics[width=7.1cm]{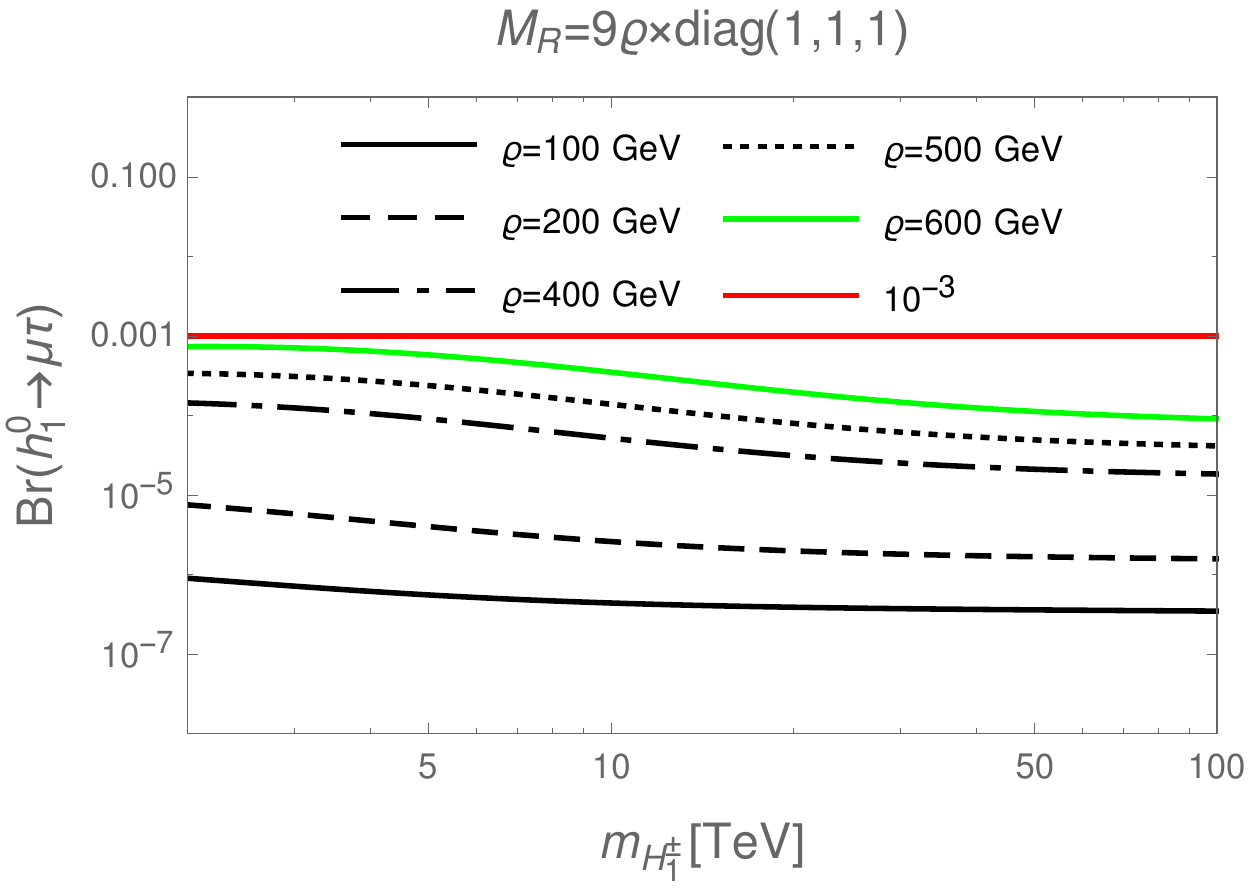} &
 		\includegraphics[width=7.0cm]{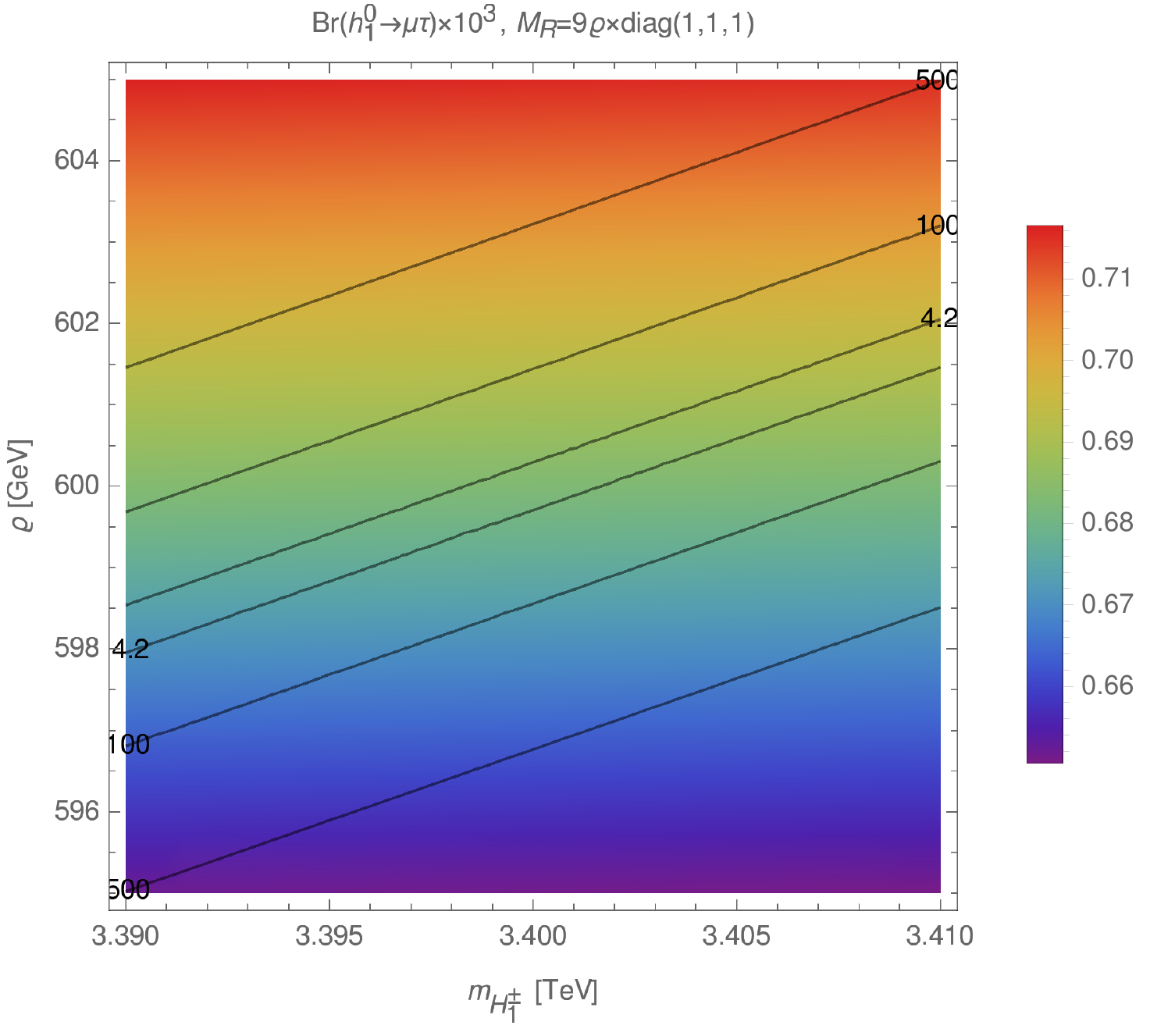} \\
 	\end{tabular}%
 	\caption{ Plots of $Br(h^0_1\rightarrow \mu\tau)$  as function of $m_{H_1^\pm}$ with fixed values $\varrho=100,200,400,500,600 \mathrm{GeV}$ and $M_R=9\varrho \times diag(1,1,1)$ (left) and density plots of $Br(h^0_1\rightarrow \mu\tau)$  as function of $m_{H_1^\pm}$ and $\varrho$ (right). The black curves in the right panel show the constant values of Br$(\mu\rightarrow e\gamma)\times 10^{13}$. } 
 	\label{fig_neutralh1}
 \end{figure}

All contributions to $Br(h^0_1\rightarrow \mu\tau)$ in case $M_R=9\varrho diag(1,1,1)$ are presented in Fig.\ref{fig_neutralh1}. Here, we chose the fixed values of $\varrho = 100, 200, 400, 500, 600 \mathrm{GeV}$ and $m_{H^\pm_1}$ in the range $500 \mathrm{GeV}$ to $100 \mathrm{TeV}$. As the result, $Br(h^0_1\rightarrow \mu\tau)$ increases with value of $\varrho$ and changes very slowly with the change of large $m_{H^\pm_1}$. The maximum value that $Br(h^0_1\rightarrow \mu\tau)$  can reach is about $0.71\times 10 ^ {-3}$. This result is very close upper limit of current experimental data, which is given in Ref.\cite{Patrignani:2016xqp}. 

We use free parameters derived from $M_R$ and $m_D$ matrices to give the results above. It is necessary to emphasize the difference with Ref.\cite{Nguyen:2018rlb} in parameterizing the matrix $m_D$, we parameterize the matrix $m_D$ in the same form as Eq.(\ref{nmD}). Using that consequence and including all contributions, especially heavy neutrinos, we have shown that $Br(h^0_1\rightarrow \mu\tau)$ is close to $10 ^ {-3}$.

A very interesting result in this model is $Br(h^0_1\rightarrow \mu\tau)$  can strongly change when choosing matrix $M_R$ with hierarchical form. To prove this statement, we investigate $Br(h^0_1\rightarrow \mu\tau)$ in cases $M_R=9\varrho \times diag(3,2,1)$ and $M_R=9\varrho \times diag(1,2,3)$. \\

\begin{figure}[ht]
	\centering
	\begin{tabular}{cc}
		\includegraphics[width=7.0cm]{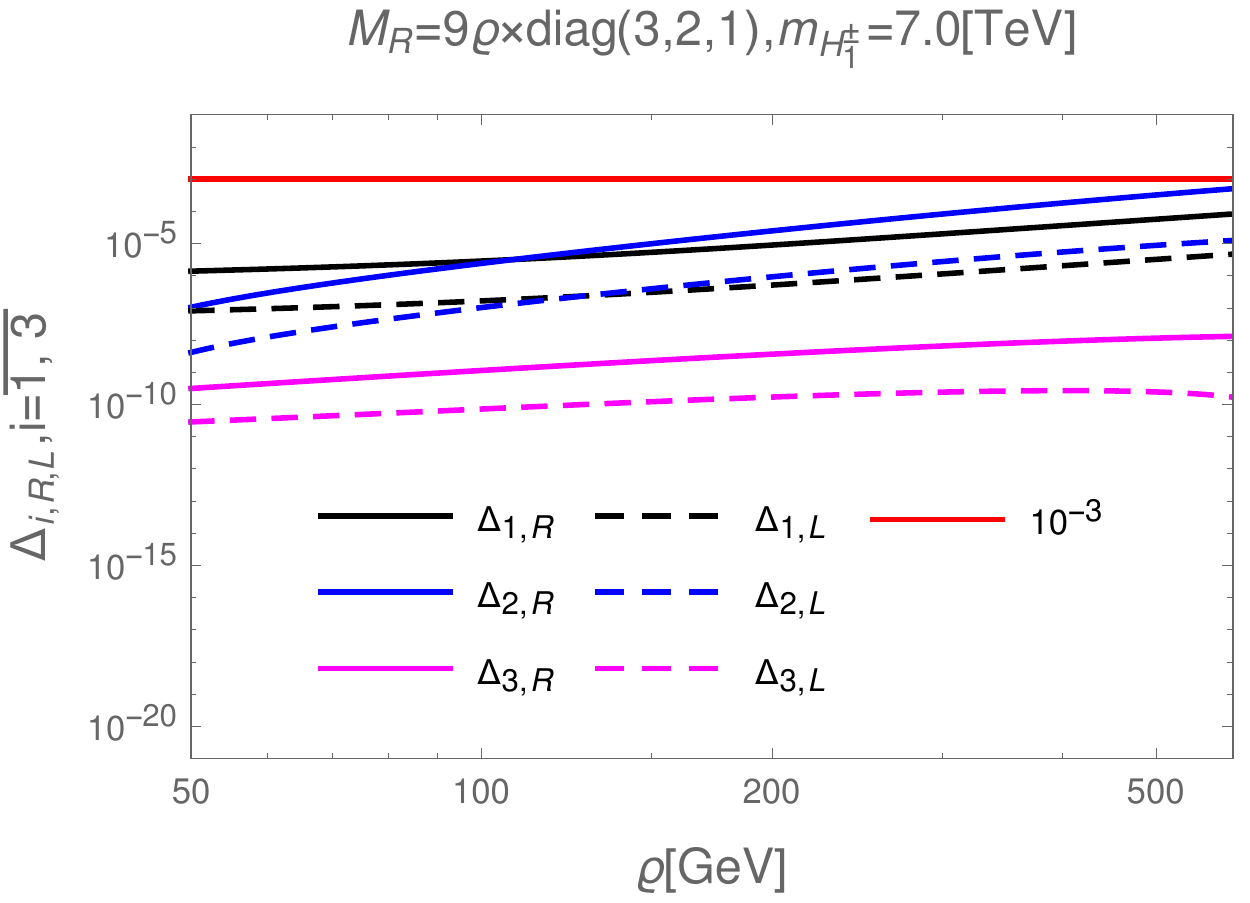} &
		\includegraphics[width=7.1cm]{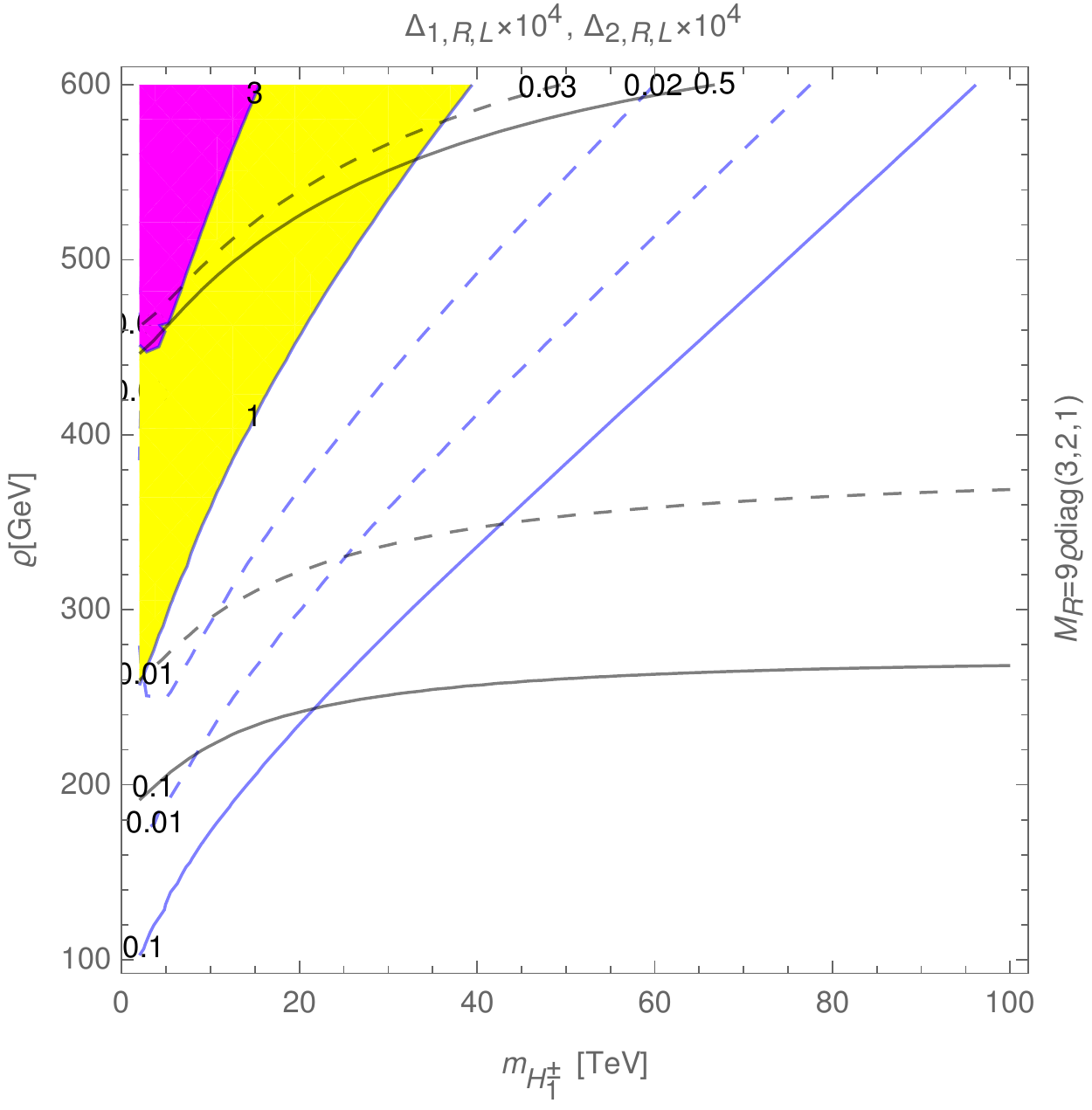} \\
	\end{tabular}%
	\caption{ Plots of $\Delta_{i,R,L},i=\overline{1,3}$  as functions of $\varrho$ with fixed value $m_{H_1^\pm}=7.0 \mathrm{TeV}$ (left) and contour plots of $\Delta_{1,2,R,L}$ as functions of $m_{H_1^\pm}$ and $\varrho$ (right). In the right panel, the black, blue, dashed-black and dashed-blue curves present constant values of $\Delta_{1,R},\Delta_{2,R},\Delta_{1,L},\Delta_{2,L}$, respectively. All plots are investigated in case $M_R=9\varrho \times diag(3,2,1)$.}
	\label{fig_con2}
\end{figure}
\begin{figure}[ht]
	\centering
	\begin{tabular}{cc}
		\includegraphics[width=7.0cm]{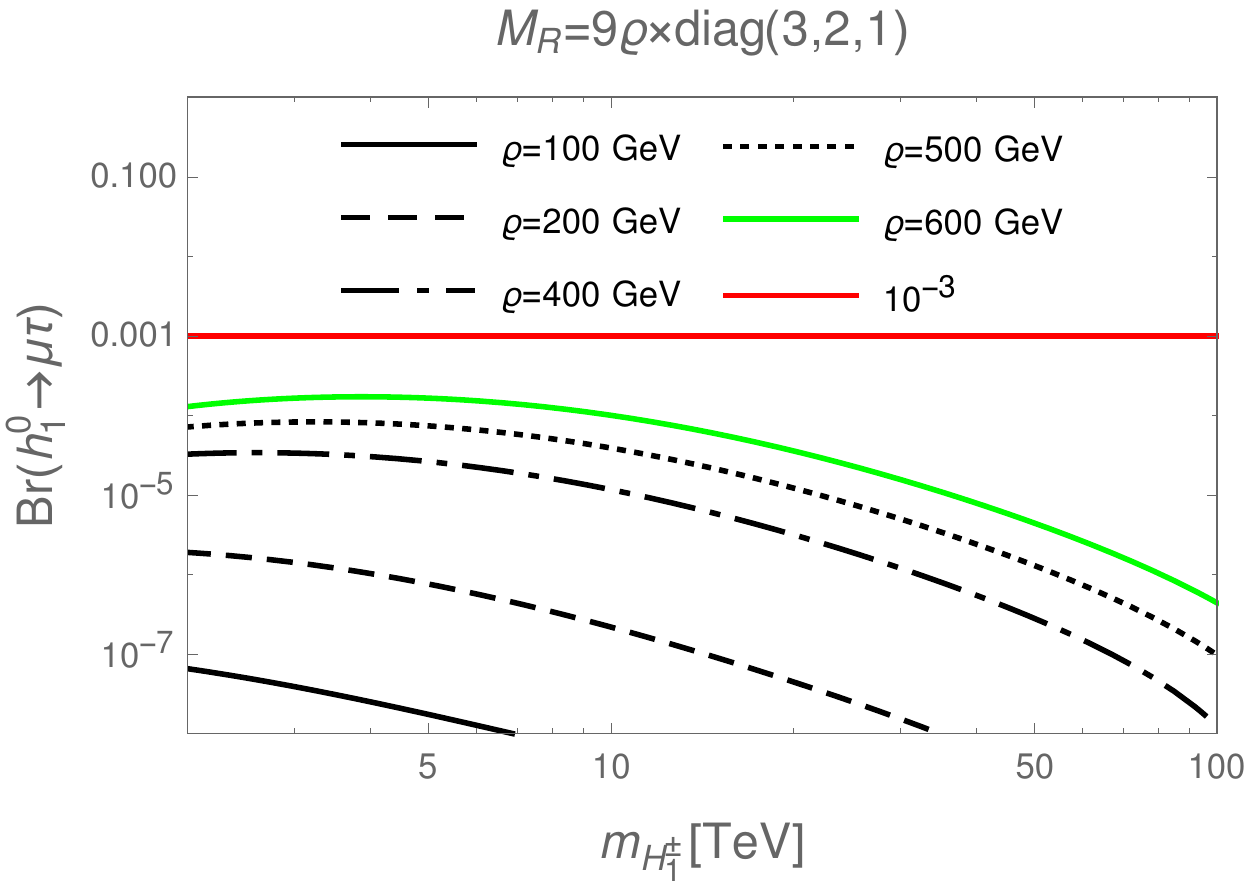} &
		\includegraphics[width=7.1cm]{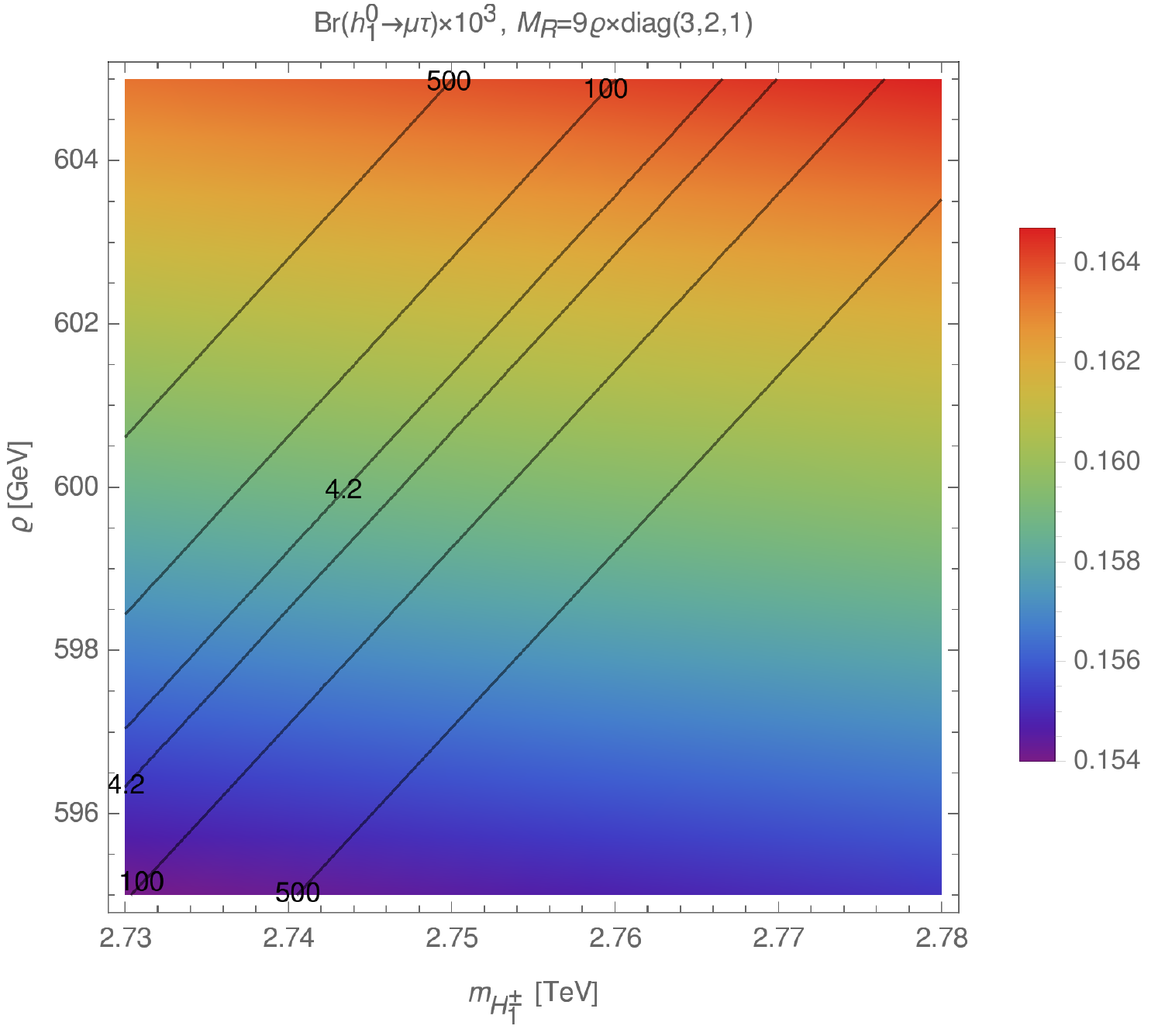} \\
	\end{tabular}%
	\caption{ Plots of $Br(h^0_1\rightarrow \mu\tau)$  as function of $m_{H_1^\pm}$ with fixed value $\varrho=100,200,400,500,600 \mathrm{GeV}$ and $M_R=9\varrho diag(3,2,1)$ (left) and density plots of $Br(h^0_1\rightarrow \mu\tau)$  as function of $m_{H_2^\pm}$ and $\varrho$ with $M_R=9\varrho \times diag(3,2,1)$ (right). The black curves in the right panel show the constant values of Br$(\mu\rightarrow e\gamma)\times 10^{13}$. }
	\label{fig_neutralh2}
\end{figure}

Ignoring contributions of $\Delta_3$ to $Br(h^0_1\rightarrow \mu\tau)$, the main contribution in case $M_R=9\varrho \times diag(3,2,1)$ is shown on the right panel of Fig.\ref{fig_con2}. The yellow, magenta present areas of $\Delta_{2,R} >1\times 10^{-4}$. It is easy to point out that the magenta region may give the largest $Br(h^0_1\rightarrow \mu\tau)$, with $\Delta_{(\mu\tau)R}\sim 3.5\times 10^{-4}$ and $\Delta_{(\mu\tau)L}\sim 0.23\times 10^{-4}$. These results lead to contributions to $Br(h^0_1\rightarrow \mu\tau)$ as shown in the left panel of Fig.\ref{fig_neutralh2}.

In the parameter space satisfying the experimental limits of $l_a \rightarrow l_b \gamma$, $Br(h^0_1\rightarrow \mu\tau)$ could reach $0.164\times 10^{-3}$ (right panel in Fig.\ref{fig_neutralh2}), smaller than the corresponding value in the case $M_R=9\varrho \times diag(1,1,1)$. 

\begin{figure}[ht]
	\centering
	\begin{tabular}{cc}
		\includegraphics[width=7.0cm]{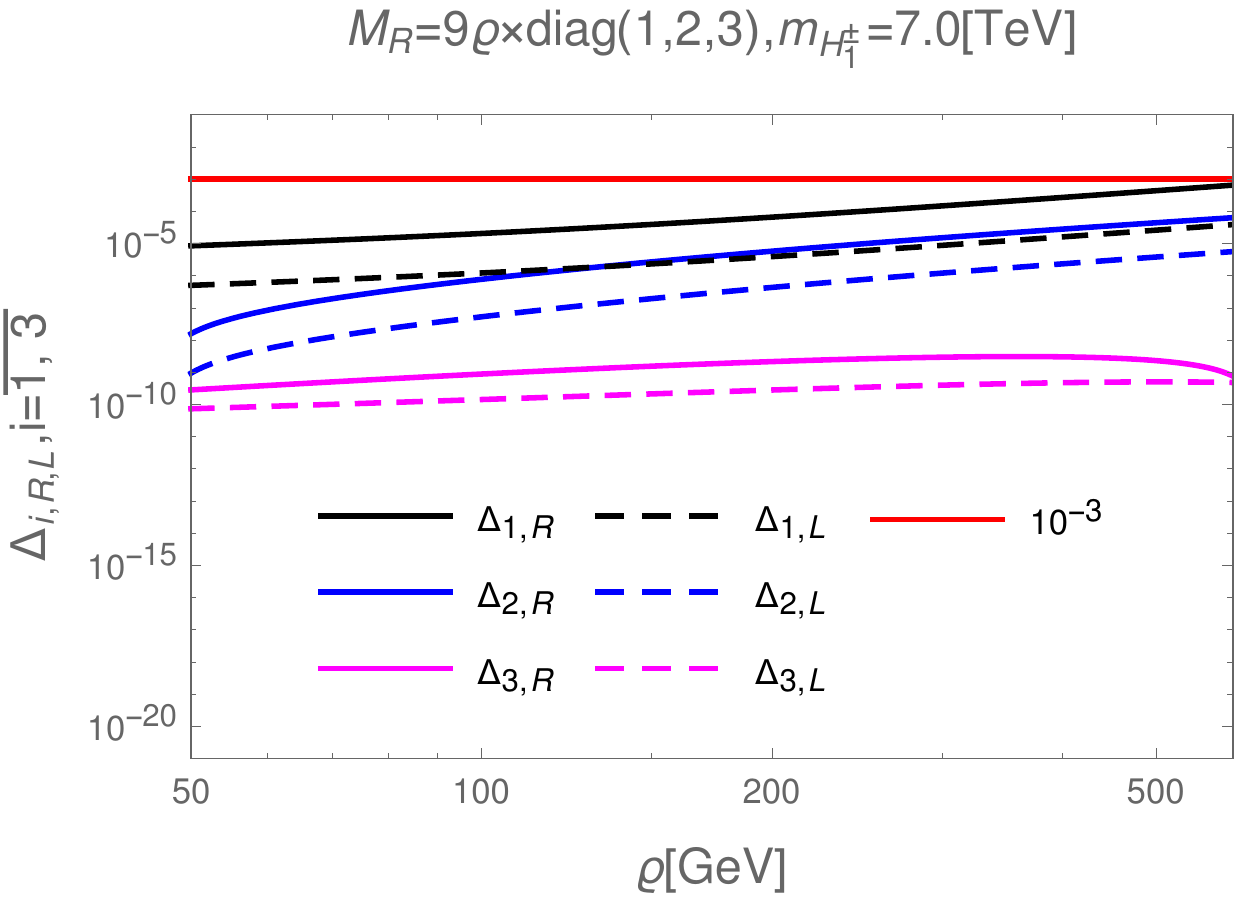} &
		\includegraphics[width=7.1cm]{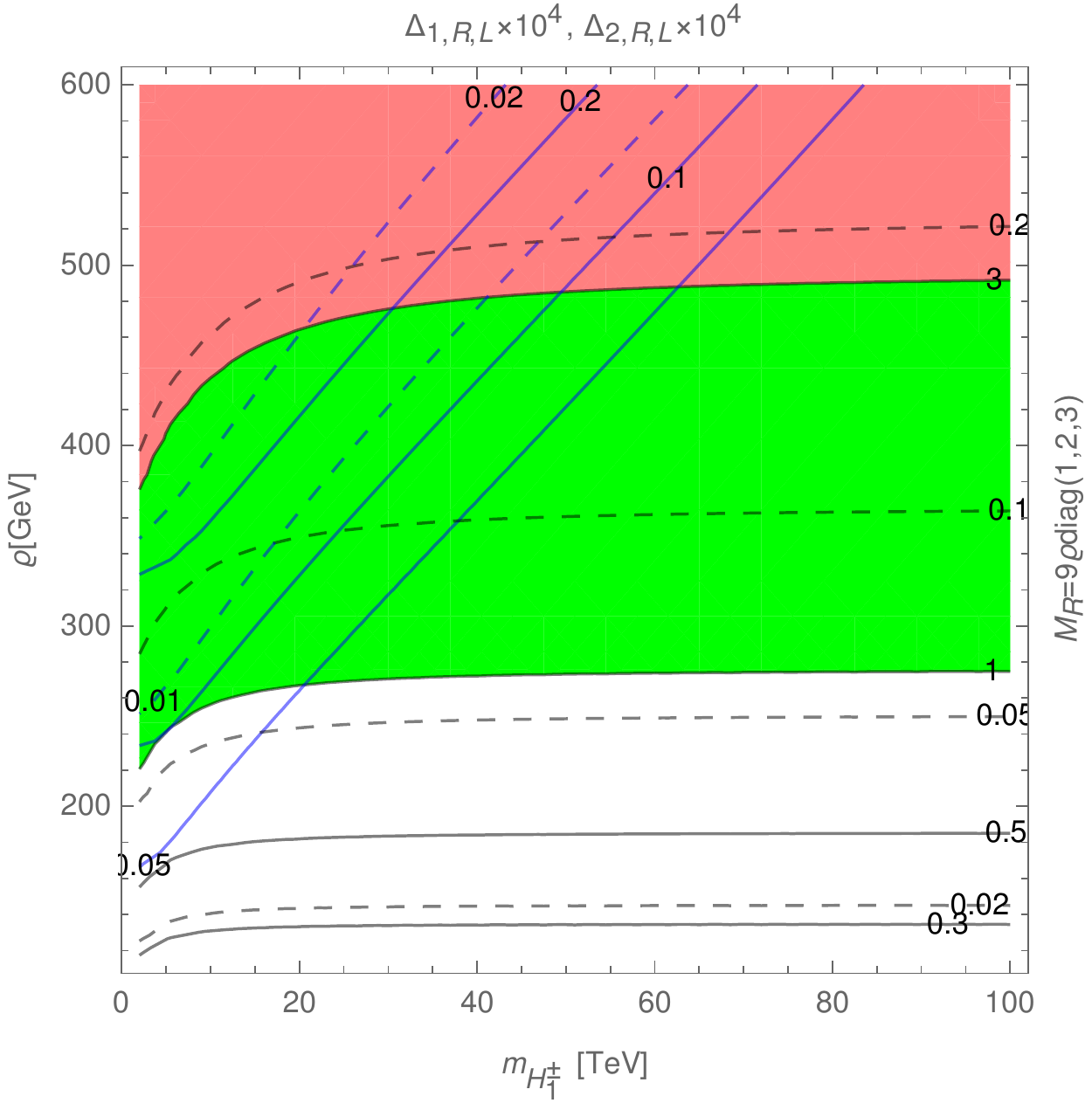} \\
	\end{tabular}%
	\caption{ Plots of $\Delta_{i,R,L},i=\overline{1,3}$  as functions of $\varrho$ with fixed value $m_{H_1^\pm}=7.0 \mathrm{TeV}$ (left) and contour plots of $\Delta_{1,2,R,L}$ as functions of $m_{H_1^\pm}$ and $\varrho$ (right). In the right panel, the black, blue, dashed-black and dashed-blue curves present constant values of $\Delta_{1,R},\Delta_{2,R},\Delta_{1,L},\Delta_{2,L}$, respectively. All plots are investigated in case $M_R=9\varrho \times diag(1,2,3)$}
	\label{fig_con3}
\end{figure}

Similarly, we can obtain the pink area in the right panel of Fig.\ref{fig_con3} that is likely to give the largest value of $Br(h^0_1\rightarrow \mu\tau)$ when $\Delta_{(\mu\tau)R}\sim 3.2\times 10^{-4}$ and $\Delta_{(\mu\tau)L}\sim 0.22\times 10^{-4}$ in case $M_R=9\varrho \times  diag(1,2,3)$.

\begin{figure}[ht]
	\centering
	\begin{tabular}{cc}
		\includegraphics[width=7.0cm]{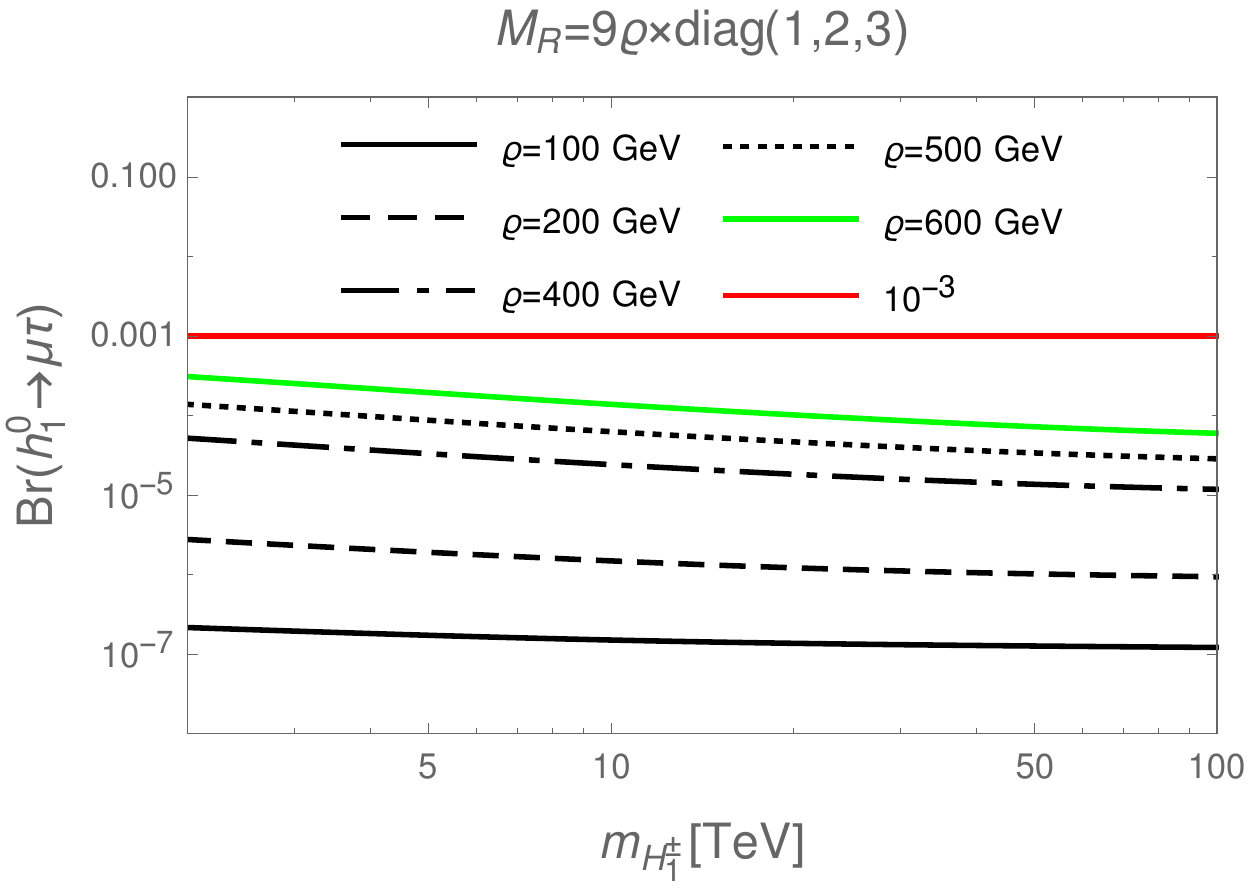} &
		\includegraphics[width=7.1cm]{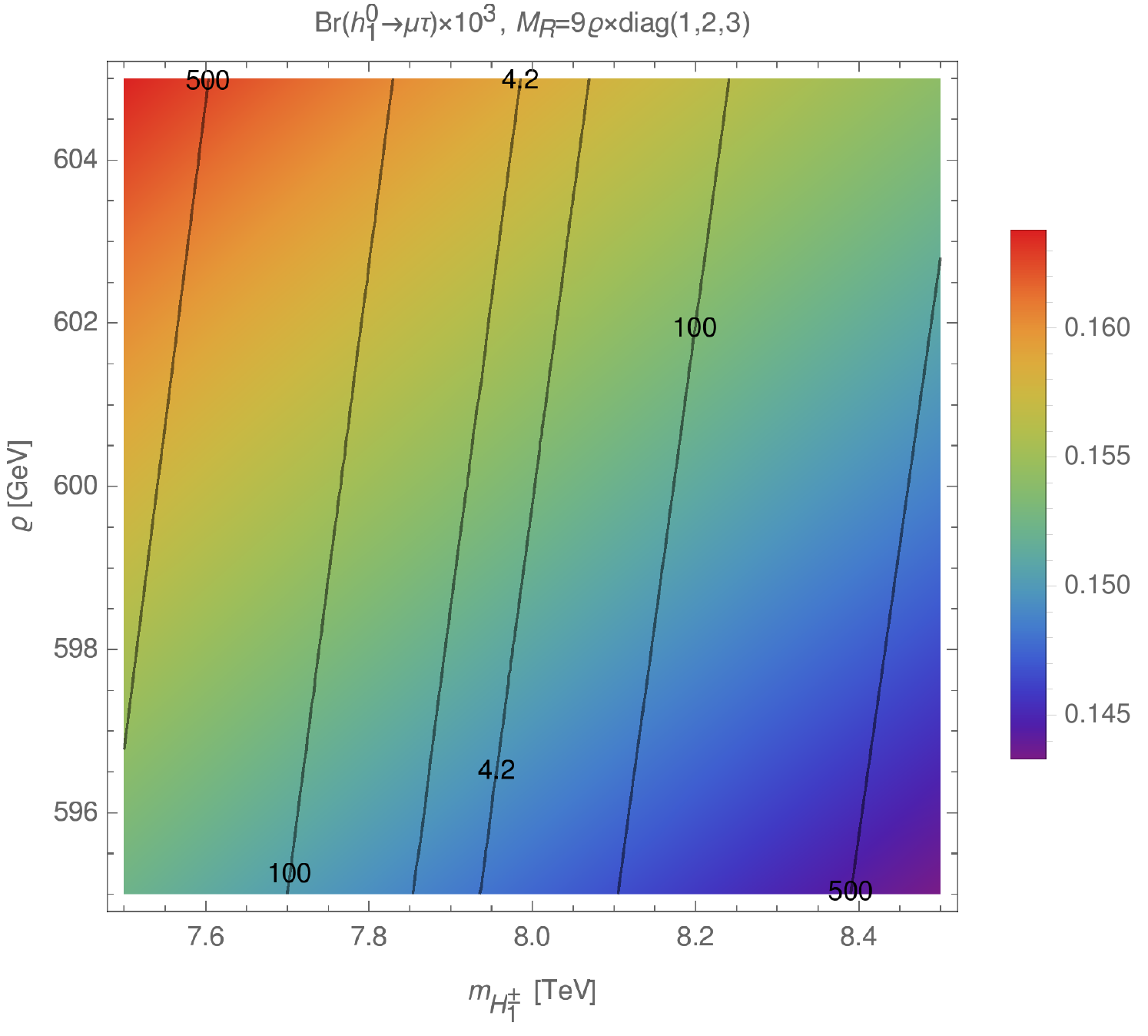} \\
	\end{tabular}%
	\caption{ Plots of $Br(h^0_1\rightarrow \mu\tau)$  as function of $m_{H_1^\pm}$ with fixed values $\varrho=100,200,400,500,600 \mathrm{GeV}$ and $M_R=9\varrho diag(1,2,3)$ (left) and density plots of $Br(h^0_1\rightarrow \mu\tau)$  as function of $m_{H_1^\pm}$ and $\varrho$ with $M_R=9\varrho \times diag(1,2,3)$ (right). The black curves in the right panel show the constant values of Br$(\mu\rightarrow e\gamma)\times 10^{13}$.}
	\label{fig_neutralh3}.
\end{figure}

The change rule of $\Delta_{i,R,L},i=\overline{1,3}$ in Fig.\ref{fig_con3} produces the survey results of $Br(h^0_1\rightarrow \mu\tau)$ as shown in Fig.\ref{fig_neutralh3}. The largest value of $Br(h^0_1\rightarrow \mu\tau)$ as presented in the right panel of Fig.\ref{fig_neutralh3} is about $0.160\times 10^{-3}$. This value is approximately to corresponding ones in case $M_R=9\varrho \times diag(3,2,1)$, but also smaller when $M_R=9\varrho \times diag(1,1,1)$.

In fact, when $M_R$ is chosen in different diagonal form, the heavy neutrinos ($N_a, F_a$) have different masses. This is caused that the contributions of charged Higgs  and gauge bosons to $Br(\mu \rightarrow e\gamma)$ will be destructive interference at the different $m_{H_1^\pm}$. Carrying out numerical investigation as in part \ref{Numerical2} , we also show that in the  areas of parameter space that satisfied the experimental limit of $Br(\mu \rightarrow e\gamma)$ then $Br(\tau \rightarrow e\gamma)$ and $Br(\tau \rightarrow \mu\gamma)$ also satisfy.  These are consequences that the narrow regions of parameter space where satisfy the experimental limits of $Br(l_a\rightarrow l_b\gamma)$ have different ranges of $m_{H_1^\pm}$ as shown in the right panels of Fig.\ref{fig_neutralh1}, Fig.\ref{fig_neutralh2}, Fig.\ref{fig_neutralh3}.   

However, the values of $Br(h^0_1\rightarrow \mu\tau)$  are only really meaningful when considered in narrow spaces that satisfy the experimental limits of $Br(l_a\rightarrow l_b\gamma)$ . These allowed spaces are confined to the two curves $4.2$ (black) on the right parts of Fig.\ref{fig_neutralh1}, Fig.\ref{fig_neutralh2} and  Fig.\ref{fig_neutralh3}. In these regions, $Br(h^0_1\rightarrow \mu\tau)$ can reach $0.71\times 10 ^ {-3}$ in case $M_R=9\varrho \times diag(1,1,1)$. This value is close to the upper bound of the experimental limit and can be detected by large accelerators to confirm the validity of this model.
 \section{\label{conclusion} Conclusion}
 In the 331ISS model, when the Marajona neutrinos ($F_a$), which are $SU(3)_L$ singlets, were added, the neutrinos were mixed and massed according to an inverse seesaw mechanism. Therefore, lepton flavor violating couplings are generated. The gauge bosons and the charged Higgs in this model make a major contribution to the $l_a\rightarrow l_b\gamma$ decays. Investigating the participation of heavy neutrinos in these major contributions, we show that these components are sometimes mutually destructive. Due to the interference of major contributions, narrow regions of the parameter space satisfying the experimental limits of the $Br(\mu\rightarrow e\gamma)$ are created and in those regions, $k$ is small and $\varrho$ is large ($k$ is the ratio factor when parameterizing the matrix $M_R$ and the matrix $m_D$ ). In particular, in these allowed narrow spaces, $Br(\tau\rightarrow e\gamma)$ can reach about $10 ^{- 9}$ and $Br(\tau\rightarrow \mu\gamma)$ may  achieve $10 ^{-10}$, these results are very close to the upper bound of the experimental limits.
 
 Performing numerical investigation, we point out that $\Delta_{1,2}$ are  the main contributions to $Br(h^0_1\rightarrow \mu\tau)$ while $\Delta_3$ is ignored because it is very small compared to $\Delta_{1,2}$. All of these contributions are less than $10 ^{- 3}$ in the selected parameter space of this model.
 
 We also found that the contributions of heavy neutrinos through $\Delta_i,\,i=\overline{1,3}$ lead to the change of $Br(h^0_1\rightarrow \mu\tau)$. This is presented through the hierarchy of the mixing matrix of heavy neutrinos ($M_R$). In case $M_R \sim diag(1,1,1)$, $Br(h^0_1\rightarrow \mu\tau)$ has a greater value than the cases $M_R \sim diag(3,2,1)$ and $M_R \sim diag(1,2,3)$.  The largest value that  $Br(h^0_1\rightarrow \mu\tau)$ can reach is about $\mathcal{O}(10 ^{-3})$ in the context of this model.  
 \section*{Acknowledgments}
 The authors would like to thanks Dr.L.T. Hue for useful discussions about applying the inverse seesaw mechanism in the 331ISS model. This research is funded by Vietnam National Foundation for Science and Technology Development (NAFOSTED) under grant number 103.01-2020.01.
 
 \appendix
 \section{\label{DeltaLR}Form factors of LFVHDs in the unitary gauge}
 In this appendix, we use Passarino-Veltman (PV) functions \cite{Hue:2015fbb,Phan:2016ouz} for representing all analytic formulas of one-loop contributions to LFVHDs defined in Eq. (\ref{LFVwidth}). We also use notations for one-loop integral of PV functions, such as $D_0=k^2-M_0^2+i\delta$, $D_1=(k-p_1)^2-M_1^2+i\delta$, $D_2=(k+p_2)^2-M_2^2+i\delta$ where $\delta$ is an infinitesimal positive real quantity.
  \bea
 B^{(i)}_{0,\mu} &\equiv&\frac{\left(2\pi\mu\right)^{4-D}}{i\pi^2}\int \frac{d^D k\left\{1,k_{\mu}\right\}}{D_0D_i},\quad
 B^{(12)}_0 \equiv \frac{\left(2\pi\mu\right)^{4-D}}{i\pi^2}\int \frac{d^D k}{D_1D_2},\crn
 C_{0,\mu} &\equiv&   C_{0,\mu}(M_0,M_1,M_2) =\frac{1}{i\pi^2}\int \frac{d^4 k\left\{1,k_{\mu}\right\}}{D_0D_1D_2},\crn
 B^{(i)}_{\mu}&=&B^{(i)}_1p_{i\mu}, \hs C_{\mu}=C_1 p_{1\mu} + C_2 p_{2\mu}.\nn
 \label{scalrInte}\eea
 The analytic expressions for $\Delta_{L,R}^{(k)W}\equiv\Delta_{(ab)L,R}^{(k)W}$, $\Delta_{L,R}^{(k)H_s}\equiv\Delta_{(ab)L,R}^{(k)H_s}$ and $\Delta_{L,R}^{(k)Y}\equiv\Delta_{(ab)L,R}^{(k)YH_1^\pm}+\Delta_{(ab)L,R}^{(k)YH_2^\pm}$ where $k$ implies the diagram (k) in Fig. \ref{fig_hmt}, are divided into the following sections.
 
 Donations with the participation of $W^\pm$ -boson
 \bea
 \Delta^{(1)W}_{L} &=& \frac{g^3c_{\beta}  m_{a}}{64\pi^2 m_W^3}\sum_{i=1}^{9}U^{\nu*}_{ai}U^{\nu}_{bi}
 \left\{ m_{n_i}^2\left(B^{(1)}_1- B^{(1)}_0- B^{(2)}_0\right) -m_b^2 B^{(2)}_1  +\left(2m_W^2+m^2_{h^0_1}\right)m_{n_i}^2 C_0 \right.\crn &-&\left. \left[2m_W^2\left(2m_W^2+m_{n_i}^2+m_a^2-m_b^2\right) + m_{n_i}^2m_{h^0_1}^2\right] C_1 +
 \left[2m_W^2\left(m_a^2-m^2_{h^0_1}\right)+ m_b^2 m^2_{h^0_1}\right]C_2\frac{}{}\right\},\crn
 \Delta^{(1)W}_{R}&=& \frac{g^3 c_{\beta}m_b}{64\pi^2 m_W^3}\sum_{i=1}^{9}U^{\nu*}_{ai}U^{\nu}_{bi}
 \left\{ -m_{n_i}^2\left(B^{(2)}_1+B^{(1)}_0+ B^{(2)}_0\right) +m_a^2 B^{(1)}_1  +\left(2m_W^2+m^2_{h^0_1}\right)m_{n_i}^2 C_0 \right.\crn &-&\left.
 \left[2m_W^2\left(m_b^2-m^2_{h}\right)+ m_a^2 m^2_{h^0_1}\right]C_1 + \left[2m_W^2\left(2m_W^2+m_{n_i}^2-m_a^2+m_b^2\right) + m_{n_i}^2m_{h^0_1}^2\right] C_2 \frac{}{}\right\},\crn
  \Delta^{(4+5)W}_{L} &=& \frac{g^3m_am_b^2c_{\beta}}{64\pi^2m^3_W(m_a^2-m_b^2)} \sum_{i=1}^{9}U^{\nu*}_{ai}U^{\nu}_{bi}\left[  2m_{n_i}^2\left(B^{(1)}_0-B^{(2)}_0\right) \right. \crn&-&\left. \left(2 m_W^2 +m_{n_i}^2\right) \left(B^{(1)}_1 +B^{(2)}_1 \right)- m_a^2 B^{(1)}_1 -m_b^2 B^{(1)}_2 \right],  
 \crn
 \Delta^{(4+5)W}_{R} &=&\frac{m_a}{m_b}\Delta^{(4+5)W}_{L}, \crn
 \Delta^{(8)W}_{L} &=& \frac{g^3c_{\beta} m_a}{64\pi^2   m_W^3}\sum_{i,j=1}^{9}U^{\nu*}_{ai}U^{\nu}_{bj} \left\{\lambda^{0*}_{ij}m_{n_j}\left[B^{(12)}_0-m_W^2C_0+\left(2 m_W^2+m_{n_i}^2-m_a^2\right)C_1\right]\right. \crn
 &&\hspace{4cm}\left.+\lambda^{0}_{ij}m_{n_i}\left[B^{(1)}_1+\left(2 m_W^2+m_{n_j}^2-m_b^2\right)C_1\right]\right\},\crn
 \Delta^{(8)W}_{R} &=&\frac{g^3 c_{\beta}m_b}{64\pi^2 m_W^3}\sum_{i=1}^{9}U^{\nu*}_{ai}U^{\nu}_{bj} \left\{\lambda^{0}_{ij}m_{n_i}\left[B^{(12)}_0-m_W^2C_0-\left(2 m_W^2+m_{n_j}^2-m_b^2\right)C_2\right]\right. \crn
 &&\hspace{3.7cm}-\left.\lambda^{0*}_{ij}m_{n_j}\left[B^{(2)}_1+\left(2 m_W^2+m_{n_i}^2-m_a^2\right)C_2\right]\right\}.
 \label{deltaLRW}\eea
 Donations with the participation of $Y^\pm$ -boson

 \bea \Delta^{(1)Y}_{L} &=&-\frac{g^3  m_{a}\left(\sqrt{2}s_{\beta}c_{\alpha}-c_{\beta}s_{\alpha}\right)}{64 \sqrt 2 \pi^2 m_Y^3}\sum_{i=1}^{9}U^{\nu*}_{(a+3)i}U^{\nu}_{(b+3)i}
 \left\{ m_{n_i}^2\left(B^{(1)}_1- B^{(1)}_0-  B^{(2)}_0\right) -m_b^2 B^{(2)}_1\right.\crn
 &+&\left.\left(2m_Y^2+m^2_{h^0_1}\right)m_{n_i}^2 C_0 -  \left[2m_Y^2\left(2m_Y^2+m_{n_i}^2+m_a^2-m_b^2\right)  + m_{n_i}^2m_{h^0_1}^2\right] C_1 \right.\crn
 &+&\left.
 \left[2m_Y^2\left(m_a^2-m^2_{h^0_1}\right)+ m_b^2 m^2_{h^0_1}\right]C_2\frac{}{}\right\},\crn
 \Delta^{(1)Y}_{R} &=& -\frac{g^3 m_{b}\left(\sqrt{2}s_{\beta}c_{\alpha}-c_{\beta}s_{\alpha}\right)}{64 \sqrt 2 \pi^2 m_Y^3}\sum_{i=1}^{9}U^{\nu*}_{(a+3)i}U^{\nu}_{(b+3)i}
 \left\{-m_{n_i}^2\left(B^{(2)}_1+B^{(1)}_0+ B^{(2)}_0\right) +m_a^2 B^{(1)}_1 \right.\crn
 &+&\left.\left(2m_Y^2+m^2_{h^0_1}\right)m_{n_i}^2 C_0-\left[2m_Y^2\left(m_b^2-m^2_{h^0_1}\right)+ m_a^2 m^2_{h^0_1}\right]C_1  \right.\crn &+&\left.
 \left[2m_Y^2\left(2m_Y^2+m_{n_i}^2-m_a^2+m_b^2\right) + m_{n_i}^2m_{h^0_1}^2\right] C_2 \frac{}{}\right\},\crn
 \Delta^{(2)Y}_{L} &=&\frac{g^3 m_{a} c_{\alpha}\left( c_\beta c_\alpha + \sqrt 2 s_\beta s_\alpha \right) }{64 \pi^2 m_Wm_Y^2}\sum_{i=1}^{9}U^{\nu*}_{(a+3)i}\crn
 &\times&\left\{\lambda^{L,2}_{bi}m_{n_i}\left[ B^{(1)}_0-B^{(1)}_1+\left(m_Y^2+m_{H^\pm_1}^2-m_{h^0_1}^2\right)C_0+\left(m_Y^2-m_{H^\pm_1}^2+m_{h^0_1}^2\right)C_1\right]\right.\crn
 && +\left. \lambda^{R,2}_{bi}m_{b}\left[ 2m_Y^2C_1-\left(m_Y^2+m_{H^\pm_1}^2-m_{h^0_1}^2\right)C_2\right]\right\},\crn
 \Delta^{(2)Y}_{R} &=&\frac{g^3 c_{\alpha}\left(c_{\beta}c_{\alpha} +\sqrt{2}s_{\beta}s_{\alpha}\right)}{64 \pi^2 m_Wm_Y^2}\sum_{i=1}^{9}U^{\nu*}_{(a+3)i}\crn
 &\times&\left\{\lambda^{L,2}_{bi}m_bm_{n_i}\left[ -2m_Y^2C_0-\left(m_Y^2-m_{H^\pm_1}^2+m_{h^0_1}^2\right)C_2\right]\right.\crn
 &&+
 \left. \lambda^{R,2}_{bi}\left[-m_{n_i}^2 B^{(1)}_0+m_a^2B^{(1)}_1+m_{n_i}^2\left(m_Y^2-m_{H^\pm_1}^2+m_{h^0_1}^2\right)C_0\right.\right.\crn
 &&+\left.\left.\left[ 2m_Y^2\left(m_{h^0_1}^2-m_b^2\right)- m_a^2\left(m_Y^2-m_{H^\pm_1}^2+m_{h^0_1}^2\right)\right]C_1 +2 m_b^2m_Y^2C_2\right]\right\},\crn
 \Delta^{(3)Y}_{L} &=&\frac{g^3 c_{\alpha}\left(c_{\beta}c_{\alpha} +\sqrt{2}s_{\beta}s_{\alpha}\right)}{64\pi^2 m_Wm_Y^2}\sum_{i=1}^{9}U^{\nu}_{(b+3)i}\crn
 &\times&\left\{\lambda^{L,2*}_{ai}m_am_{n_i}\left[ -2m_Y^2C_0+\left(m_Y^2-m_{H^\pm_1}^2+m_{h^0_1}^2\right)C_1\right]\right.\crn
 &&+
 \left. \lambda^{R,2*}_{ai}\left[-m_{n_i}^2 B^{(2)}_0-m_b^2B^{(2)}_1+m_{n_i}^2\left(m_Y^2-m_{H^\pm_1}^2+m_{h^0_1}^2\right)C_0\right.\right.\crn
 &&-\left.\left. 2m_a^2m_Y^2C_1-\left[ 2m_Y^2\left(m_{h^0_1}^2-m_a^2\right)- m_b^2\left(m_Y^2-m_{H^\pm_1}^2+m_{h^0_1}^2\right)\right]C_2\right]\right\},\crn
 \Delta^{(3)Y}_{R} &=&\frac{g^3 m_b c_{\alpha}\left(c_{\beta}c_{\alpha} +\sqrt{2}s_{\beta}s_{\alpha}\right)}{64\pi^2 m_Wm_Y^2}\sum_{i=1}^{9}U^{\nu}_{(b+3)i}\crn
 &\times&\left\{\lambda^{L,2*}_{ai}m_{n_i}\left[ B^{(2)}_0+B^{(2)}_1+\left(m_Y^2+m_{H^\pm_1}^2-m_{h^0_1}^2\right)C_0-\left(m_Y^2-m_{H^\pm_1}^2+m_{h^0_1}^2\right)C_2\right]\right.\crn
 && +\left. \lambda^{R,2*}_{ai}m_{a}\left[  \left(m_Y^2+m_{H^\pm_1}^2-m_{h^0_1}^2\right)C_1-2m_Y^2C_2\right]\right\},\crn
 \Delta^{(4+5)Y}_{L} &=&\frac{g^3m_am_b^2c_{\beta}}{64\pi^2m_Wm_Y^2(m_a^2-m_b^2)} \sum_{i=1}^{9}U^{\nu*}_{(a+3)i}U^{\nu}_{(b+3)i} \crn &\times&\left[  2m_{n_i}^2\left(B^{(1)}_0-B^{(2)}_0\right)-\left(2 m_Y^2 +m_{n_i}^2\right) \left(B^{(1)}_1 +B^{(2)}_1 \right) - m_a^2 B^{(1)}_1 -m_b^2 B^{(2)}_1 \right], \crn
 \Delta^{(4+5)Y}_{R} &=&\frac{m_a}{m_b}\Delta^{(4+5)Y}_{L}, \crn
 \Delta^{(8)Y}_{L} &=&\frac{g^3c_{\beta} m_a}{64\pi^2  m_Wm_Y^2}\crn
 &\times&\sum_{i,j=1}^{9}U^{\nu*}_{(a+3)i}U^{\nu}_{(b+3)j} \left\{\lambda^{0*}_{ij}m_{n_j}\left[B^{(12)}_0-m_Y^2C_0+\left(2 m_Y^2+m_{n_i}^2-m_a^2\right)C_1\right]\right. \crn
 &&\hspace{3.5cm}+\left.\lambda^{0}_{ij}m_{n_i}\left[B^{(1)}_1+\left(2 m_Y^2+m_{n_j}^2-m_b^2\right)C_1\right]\right\},\crn
 \Delta^{(8)Y}_{R} &=&\frac{g^3c_{\beta} m_b}{64\pi^2 m_Wm_Y^2}\crn
 &\times&\sum_{i,j=1}^{9}U^{\nu*}_{(a+3)i}U^{\nu}_{(b+3)j} \left\{\lambda^{0}_{ij}m_{n_i}\left[B^{(12)}_0-m_Y^2C_0-\left(2 m_Y^2+m_{n_j}^2-m_b^2\right)C_2\right]\right. \crn
 &&\hspace{3.cm}-\left.\lambda^{0*}_{ij}m_{n_j}\left[B^{(2)}_1+\left(2 m_Y^2+m_{n_i}^2-m_a^2\right)C_2\right]\right\}.
  \label{deltaLRY}\eea
 Donations with the participation of $H_s^\pm$ -boson
 
 \bea
 \Delta^{(6)H_s}_{L} &=&   -\frac{g^3c_{\beta}f_s}{32\pi^2  m_W^3}\sum_{i,j=1}^{9}\left\{\lambda^{0*}_{ij}\left[\lambda^{R,s*}_{ai}\lambda^{L,s}_{bj}\left(B^{(12)}_0+m_{H^\pm_s}^2C_0 -m_a^2 C_1+m_b^2C_2\right)\right.\right.\crn
 &+&\left.\left. \lambda^{R,s*}_{ai}\lambda^{R,s}_{bj}m_bm_{n_j}C_2 -\lambda^{L,s*}_{ai}\lambda^{L,s}_{bj}m_am_{n_i}C_1 \right]\right. \crn
 &+&\left.  \lambda^{0}_{ij}\left[\lambda^{R,s*}_{ai}\lambda^{L,s}_{bj}m_{n_i}m_{n_j}C_0 +\lambda^{R,s*}_{ai}\lambda^{R,s}_{bj}m_{n_i}m_{b}(C_0+C_2)\right.\right.\crn
 &+&\left.\left.\lambda^{L,s*}_{ai}\lambda^{L,s}_{bj}m_{a}m_{n_j}(C_0-C_1)+ \lambda^{L,s*}_{ai}\lambda^{R,s}_{bj}m_{a}m_{b}(C_0-C_1+C_2) \right]\frac{}{}\right\},\crn
 \Delta^{(6)H_s}_{R} &=&-\frac{g^3c_{\beta}f_s}{32\pi^2  m_W^3} \sum_{i,j=1}^{9}\left\{\lambda^{0}_{ij}\left[\lambda^{L,s*}_{ai}\lambda^{R,s}_{bj}\left(B^{(12)}_0+m_{H^\pm_s}^2C_0 -m_a^2 C_1+m_b^2C_2\right)\right.\right.\crn
 &+&\left.\left.\lambda^{L,s*}_{ai}\lambda^{L,s}_{bj}m_bm_{n_j}C_2-  \lambda^{R,s*}_{ai}\lambda^{R,s}_{bj}m_am_{n_i}C_1 \right]\right.\crn
 &+&\left.
 \lambda^{0*}_{ij}\left[\lambda^{L,s*}_{ai}\lambda^{R,s}_{bj}m_{n_i}m_{n_j}C_0 +\lambda^{L,s*}_{ai}\lambda^{L,s}_{bj}m_{n_i}m_{b}(C_0+C_2)\right.\right.\crn
 &+&\left.\left.\lambda^{R,s*}_{ai}\lambda^{R,s}_{bj}m_{a}m_{n_j}(C_0-C_1)+ \lambda^{R,s*}_{ai}\lambda^{L,s}_{bj}m_{a}m_{b}(C_0-C_1+C_2) \right]\frac{}{}\right\},\crn
 \Delta^{(7)H_{s}}_{L}&=&\frac{g^2\lambda^{\pm}_{H_s}f_s}{16\pi^2  m_W^2}\sum_{i=1}^{9}\left[-\lambda^{R,s*}_{ai}\lambda^{L,s}_{bi}m_{n_i}C_0 -\lambda^{L,s*}_{ai}\lambda^{L,s}_{bi}m_{a}C_1 +\lambda^{R,s*}_{ai}\lambda^{R,s}_{bi}m_{b}C_2 \right],\crn
 \Delta^{(7)H_{s}}_{R} &=&\frac{g^2\lambda^{\pm}_{H_s}f_s}{16\pi^2 m_W^2}\sum_{i=1}^{9}\left[-\lambda^{L,s*}_{ai}\lambda^{R,s}_{bi}m_{n_i}C_0 -\lambda^{R,s*}_{ai}\lambda^{R,s}_{bi}m_{a}C_1 +\lambda^{L,s*}_{ai}\lambda^{L,s}_{bi}m_{b}C_2 \right],\crn
 \Delta^{(9+10)H_s}_{L}  &=&-\frac{g^3c_{\beta}f_s}{32\pi^2m_W^3\left(m_a^2-m_b^2\right)} \crn
 &\times&\sum_{i=1}^{9}\left[ m_am_bm_{n_i}  \lambda^{L,s*}_{ai}\lambda^{R,s}_{bi}\left(B^{(1)}_0-B^{(2)}_0\right)+ m_{n_i}
 \lambda^{R,s*}_{ai}\lambda^{L,s}_{bi}\left(m^2_bB^{(1)}_0-m^2_aB^{(2)}_0\right)\right.\crn
 &&\left.+ m_{a}m_b \left(\lambda^{L,s*}_{ai}\lambda^{L,s}_{bi}m_b + \lambda^{R,s*}_{ai}\lambda^{R,s}_{bi}m_a\right)\left(B^{(1)}_1+ B^{(2)}_1\right)\right],\crn
 \Delta^{(9+10)H_s}_{R}  &=&-\frac{g^3c_{\beta}f_s}{32\pi^2m_W^3\left(m_a^2-m_b^2\right)}\crn
 &\times&\sum_{i=1}^{9}\left[ m_am_bm_{n_i}  \lambda^{R,s*}_{ai}\lambda^{L,s}_{bi}\left(B^{(1)}_0-B^{(2)}_0\right)+ m_{n_i}
 \lambda^{L,s*}_{ai}\lambda^{R,s}_{bi}\left(m^2_bB^{(1)}_0-m^2_aB^{(2)}_0\right)\right.\crn
 &&\left.+ m_{a}m_b \left(\lambda^{R,s*}_{ai}\lambda^{R,s}_{bi}m_b + \lambda^{L,s*}_{ai}\lambda^{L,s}_{bi}m_a\right)\left(B^{(1)}_1+ B^{(2)}_1\right)\right].
 \label{deltaLRHs} \eea 
 \section{\label{CaDV} The divergent cancellation in amplitudes}
 
 The divergent parts in terms as shown in App.\ref{DeltaLR} only contain $B$ functions, we note:  div$B^{(1)}_0=$div$B^{(2)}_0=$div$B^{(12)}_0=2$div$B^{(1)}_1=-2$ div$B^{(2)}_1=\Delta_{\epsilon}$. Ignoring the common factor of $g^3/(64\pi^2m_W^3)$ and using $1/m_Y=\sqrt{2}s_{\al}/m_W$, the divergent parts of $\Delta_L$ derived from Eq. (\ref{deltaLRW}, \ref{deltaLRY}, \ref{deltaLRHs}) are
 \bea \mathrm{div}\left[\Delta^{(1)W}_L\right]&=&m_a\Delta_{\epsilon}\times \left( -\frac{3c_{\beta}}{2}\right) \sum_{i=1}^9 U^{\nu*}_{ai}U^{\nu}_{bi} m^2_{n_i},\crn
 \mathrm{div}\left[\Delta^{(8)W}_L\right]&=&m_a \Delta_{\epsilon}\times c_{\beta}\sum_{i,j=1}^9 U^{\nu*}_{ai}U^{\nu}_{bj}\left(\lambda^{0*}_{ij} m_{n_j}+\frac{1}{2}\lambda^{0}_{ij} m_{n_i}\right),\crn
 \mathrm{div}\left[\Delta^{(4+5)W}_L\right]&=&\mathrm{div}\left[\Delta^{(4)Y}_L\right]= \mathrm{div}\left[\Delta^{(4+5)Y}_L\right]=0,\crn
 \mathrm{div}\left[\Delta^{(1)Y}_L\right]&=&m_a \Delta_{\epsilon}\times 3s_{\alpha}^3 \left(\sqrt 2 s_\beta c_\al - c_\beta s_\al \right) \sum_{i=1}^9 U^{\nu*}_{(a+3)i}U^{\nu}_{(b+3)i} m^2_{n_i},\crn
 \mathrm{div}\left[\Delta^{(2)Y}_L\right]&=&m_a\Delta_{\epsilon}\times s_{\al}^2 c_\al \left( c_\beta c_\al + \sqrt 2 s_\beta s_\al \right)\sum_{i=1}^9 U^{\nu*}_{(a+3)i}\lambda^{L,1}_{bi} m_{n_i},\crn
 \mathrm{div}\left[\Delta^{(3)Y}_L\right]&=&m_a\Delta_{\epsilon}\times  \left[-2s_{\al}^2 c_\al  \left( c_\beta c_\al + \sqrt 2 s_\beta s_\al \right)\right] \sum_{i=1}^9 U^{\nu*}_{(a+3)i}U^{\nu}_{(b+3)i} m^2_{n_i},\crn
 \mathrm{div}\left[\Delta^{(8)Y}_L\right]&=&m_a \Delta_{\epsilon}\times  2s_{\al}^2c_{\beta} \sum_{i,j=1}^9 U^{\nu*}_{(a+3)i}U^{\nu}_{(b+3)j} \left(\lambda^{0*}_{ij} m_{n_j}+\frac{1}{2}\lambda^{0}_{ij} m_{n_i}\right),\crn
 \mathrm{div}\left[\Delta^{(6)H_1^\pm}_L\right]&=&m_a\Delta_{\epsilon}\times \left(-2c_{\beta}c_{\al}^2 \right) \sum_{i,j=1}^9 U^{\nu*}_{(a+3)i}\lambda^{0*}_{ij}\lambda^{L,1}_{bj},\crn
 \mathrm{div}\left[\Delta^{(6)H_2^\pm}_L\right]&=&m_a\Delta_{\epsilon}\times \left(-c_{\beta}\right) \sum_{i,j=1}^9 U^{\nu*}_{ai}\lambda^{0*}_{ij}\lambda^{L,2}_{bj},\crn
 \mathrm{div}\left[\Delta^{(9+10)H_1^\pm}_L\right]&=&m_a\Delta_{\epsilon}\times \left( 2c_{\beta}c_{\al}^2\right) \sum_{i=1}^9 U^{\nu*}_{(a+3)i}\lambda^{L,1}_{bi} m_{n_i}, \crn
 \mathrm{div}\left[\Delta^{(9+10)H_2^\pm}_L\right]&=&m_a\Delta_{\epsilon}\times  c_{\beta} \sum_{i=1}^9 U^{\nu*}_{ai}\lambda^{L,2}_{bi} m_{n_i}.
 \label{divdeli}\eea
Similarly, the divergences of the  $\Delta^{(k)W,Y,H_s^\pm}_R$ are shown.
Using the equalities $M^{\nu}=U^{\nu*}\hat{M^{\nu}}U^{\nu\dagger}$, we can prove that
\bea \mathrm{div}\left[ \Delta_{1,L,R}\right] &=&\mathrm{div}\left[\Delta^{(1)W}_{L,R}+\Delta^{(8)W}_{L,R}+\Delta^{(6)H_1^\pm}_{L,R}+\Delta^{(9+10)H_1^\pm}_{L,R}\right] =0,
\crn
\mathrm{div}\left[ \Delta_{2,L,R}\right]&=&\mathrm{div}\left[\Delta^{(1)Y}_{L,R}+\Delta^{(2)Y}_{L,R}+\Delta^{(3)Y}_{L,R}+\Delta^{(8)Y}_{L,R}+\Delta^{(6)H_2^\pm}_{L,R}+\Delta^{(9+10)H_2^\pm}_{L,R}\right] =0,\crn
\mathrm{div}\left[ \Delta_{3,L,R}\right]&=&\mathrm{div}\left[\Delta^{(7)H_1}_{L,R}+\Delta^{(7)H_2}_{L,R}+\Delta^{(4+5)W}_{L,R}+\Delta^{(4+5)Y}_{L,R}\right] =0.
\label{usex}\eea
 \bibliographystyle{h-physrev}
 \bibliography{mainH}
\end{document}